\documentclass[a4paper]{article}
\usepackage{graphicx}

\begin{document}

\author{A. Akindinov$^1$, \and V. Golovin$^2$, \and A. Martemiyanov$^1$,
\and V. Petrov$^2$, \and A. Smirnitskiy$^1$, \and 
K. Voloshin$^{1,}$\thanks{Corresponding author. E-mail: Kirill.Voloshin@itep.ru}}
\date{}
\title{Dielectric Resistive Plate Chamber\\
as a detector for time-of-flight measurements}

\maketitle

\begin{center}
\noindent\small\em $^1$Institute for Theoretical and Experimental Physics (ITEP),\\
B.~Cheremushkinskaya 25, Moscow, 117218, Russia. \\
$^2$Center of Perspective Technologies and Apparatus (CPTA), \\
Preobrazhenskaya pl.~6/8, Moscow, 107076, Russia.
\end{center}

\begin{abstract}
Principles of operation, construction and first test results of a 
Dielectric Resistive Plate Chamber (DRPC) are described. The 
detector has shown stability of operation in the avalanche mode 
of gas amplfication within a wide range of applied voltages. 
Double-gap DRPCs have demonstrated the MIP registration efficiency
of 97\% and the time resolution of 180--200~ps. No changes in DRPC 
operation have been observed with test beam intensities up to
10$^3$~Hz/cm$^2$.
\end{abstract}

\section{Principles of operation and construction}

The first beam test results for a new 
time-of-flight (TOF) gaseous detector~--- Dielectric Resistive 
Plate Chamber (DRPC)~\cite{patent} --- were obtained in the 
slopes of R\&D program for TOF gaseous detectors intended to be 
used in the proposed nuclear experiment ALICE at LHC. 

DRPC was developed as a frontier detector between PPC and RPC.
The basic idea is to avoid large energy resolution in spark 
cases (RPC feature), keeping high time resolution altogether 
with fast response provided by PPC \cite{ppc}. A dielectric 
layer introduced between the electrodes is covered with a 
surface resistivity on the gas side which lets, in case of a 
breakdown, to decrease or limit the spark energy by charging an 
elementary plate condensor, which then discharges through the 
resistive surface. The detector has shown stability of operation 
in the avalanche mode within a wide range of applied voltages. 
Time response and energy resolution in spark events may be 
adjusted by varying the main electric parameters of the detector: 
the dielectric layer capacity and the surface resistivity.  

\begin{figure}
\begin{center}
\includegraphics*[width=.7\linewidth]{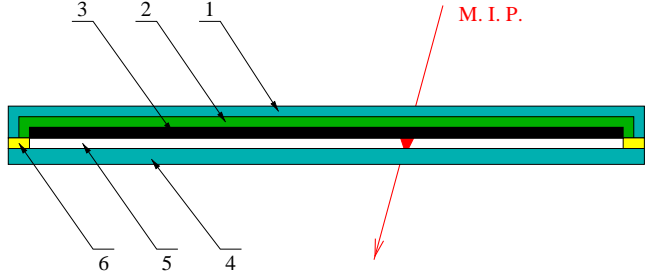}
\caption{DRPC construction: 1 ---~conducting layer (anode), 2 ---~dielectric,
  3 ---~resistive layer, 4 ---~cathode, 5 ---~gas gap, 6
  ---~spacers.} 
\label{scheme}
\end{center}
\end{figure}

The basic detector construction used in the tests is shown in
Fig.~\ref{scheme}. A plate condensor has external dimensions 
$50 \times 50$~mm$^2$, corresponding to the required dimensions 
of a single detecting cell in the ALICE TOF system, and is several 
milimeters thick. One of the electrodes is made of a 
conducting material, the other has a resistance-dielectric-metal 
structure. The gas gap has been chosen to be 0.6~mm wide basing on 
the PPC experience as this value provides high MIP registration efficiency 
altogether with a rather good time resolution. The gap is formed 
by means of ceramic spacers. Accuracy requirements are similar
to those for the traditional PPC: the electrodes and spacers 
are kept flat and parallel within the several micrometers of
accuracy. 

The layer electrode is the crucial part of the detector. The
dielectric thickness $d_d$ is determined from the condition 
$d_d \leq \epsilon_d \cdot d_g / \epsilon_g$, 
where $d_g$ is the gas gap width, 
$\epsilon_d$ and $\epsilon_g$ are the dielectric and gas
dielectric constants correspondently. We used 1~mm thick 
Al$_2$O$_3$  with
$\epsilon / \epsilon_{air} \approx 10$. The dielectric was covered
with a thin layer of resistant material SiC or TiC with the volume
resistivity $\rho = 10^2 -10^3$~$\Omega \cdot \mbox{cm}$. 
The outer side of this electrode was metallized through 
evaporation, and actually appeared to be the second condensor 
electrode producing the electric field responsible for the gas 
amplification. 

In the tests double-gap detectors formed of two identical single 
gaps were investigated. 

\begin{figure}
\begin{center}
\includegraphics*[width=.7\linewidth]{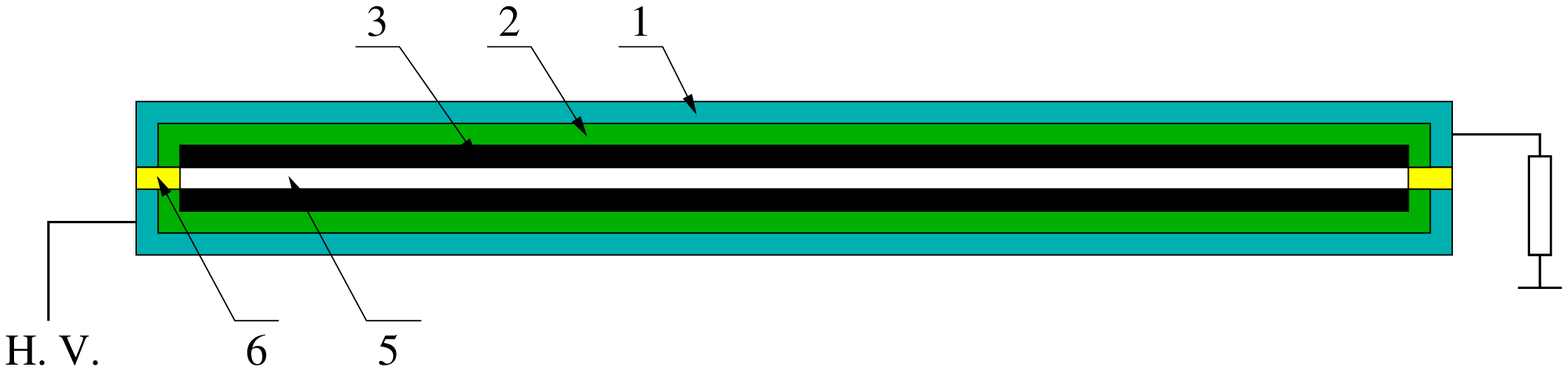}
\vspace{3mm} \\
\includegraphics[width=.7\linewidth]{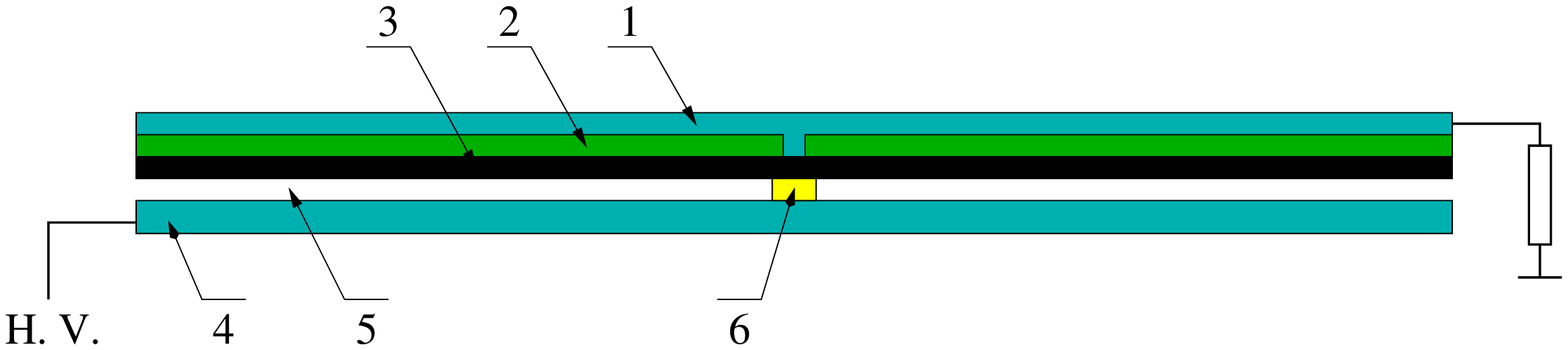}
\vspace{3mm} \\
\includegraphics[width=.7\linewidth]{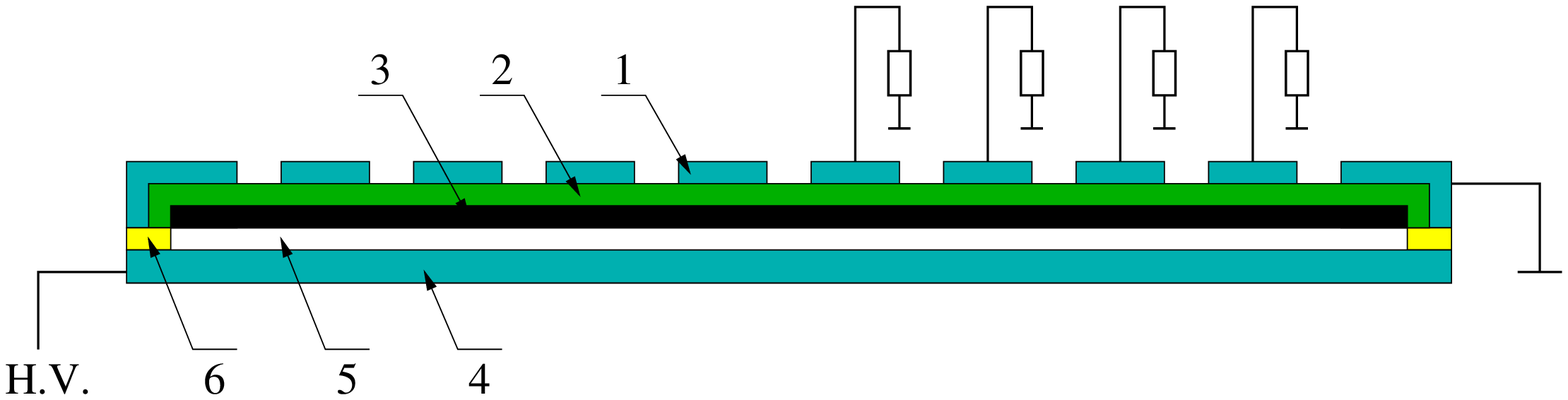}
\vspace{3mm} \\
\includegraphics[width=.7\linewidth]{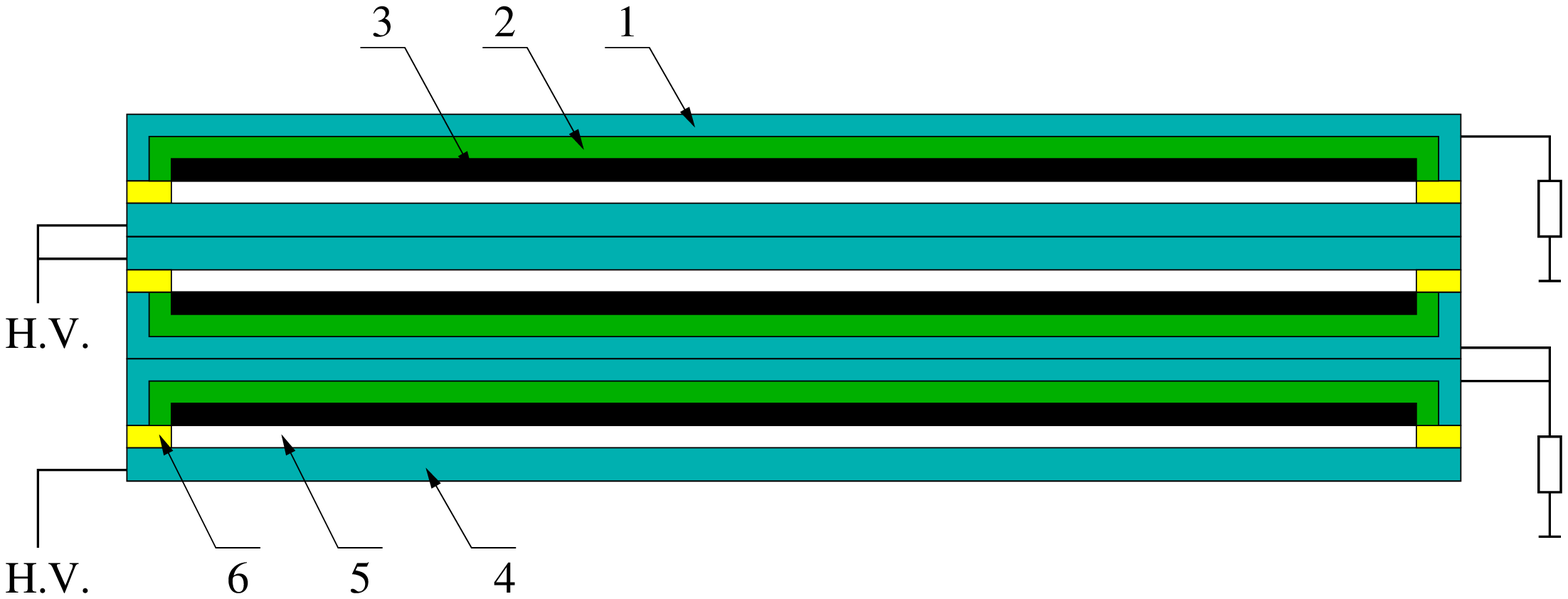}
\caption{Various options of DRPC construction: 1 ---~conducting layer (anode),
  2 ---~dielectric, 3 ---~resistive layer, 4 ---~cathode,
  5 ---~gas gap, 6 ---~spacers.} 
\label{options}
\end{center}
\end{figure}

Fig.~\ref{options} shows other possible options of the detector
assembly. 

\section{Time measurements with DRPC}

DRPC has been studied as a detector for high precision 
and efficient time measurements on PS beams at ITEP and CERN. 
$\pi^-$-beams with 2--7~GeV/$c$ momenta, 10$^2$--10$^5$~s$^{-1}$ 
intensities, and several cm$^2$ cross-section at the detector 
position were usually used. 

Trigger and start counters, made of scintillators of different 
sizes and modifications, and adapted for time measurements, were 
positioned on the same beam line. The trigger consisted of 6 to 10 
counters, including those situated on precise movable platforms, 
which determined the space position of the beam ahead and behind 
the target, important for efficiency measurements, as well as 
the crossing point between the beam and the detector plane. The 
time measurement scheme included 2 start scintillation counters 
(each with 2 PMTs), it was always possible to measure their time 
resolution continiously during the data taking. A
typical resolution of the start system equaled to 150~ps. 

The signal from the detector was used as the {\it stop} in the
scheme. The resolution of start counters was quadratically 
subtracted from the time of particles flight along a fixed base. 
In this way the detector resolution was measured. 

All trigger and timing electronics were assembled in 
the NIM and CAMAC standards. 

Specially designed fast and low-noise electronics, consisting of
a preamplifier, a main amplifier and a discriminator, were used for
registration of signals from the gaseous detector. The scheme 
comes as a development of ideas previously described
in~\cite{electronics} and lies beyond the slopes of this paper. 

Various gas mixtures, based on either DME or different types of 
freons with normal quality, were used as working gases in the 
detector. A simple gas system allowed to mix gases at rather 
small fluxes of about several litres per hour. No special 
control of quality and other gas parameters was undertaken. 

The mixture DME + 5\%CF$_3$Br was most thoroughly studied 
as a working gas for DRPC. The best results were obtained with 
C$_2$H$_2$F$_4$ + 15\%DME and C$_2$H$_2$F$_4$ + 
5\%iso\-butane + 10\%SF$_6$. 

\begin{figure}
\begin{center}
\begin{tabular}{cc}
\includegraphics[width=.45\linewidth]{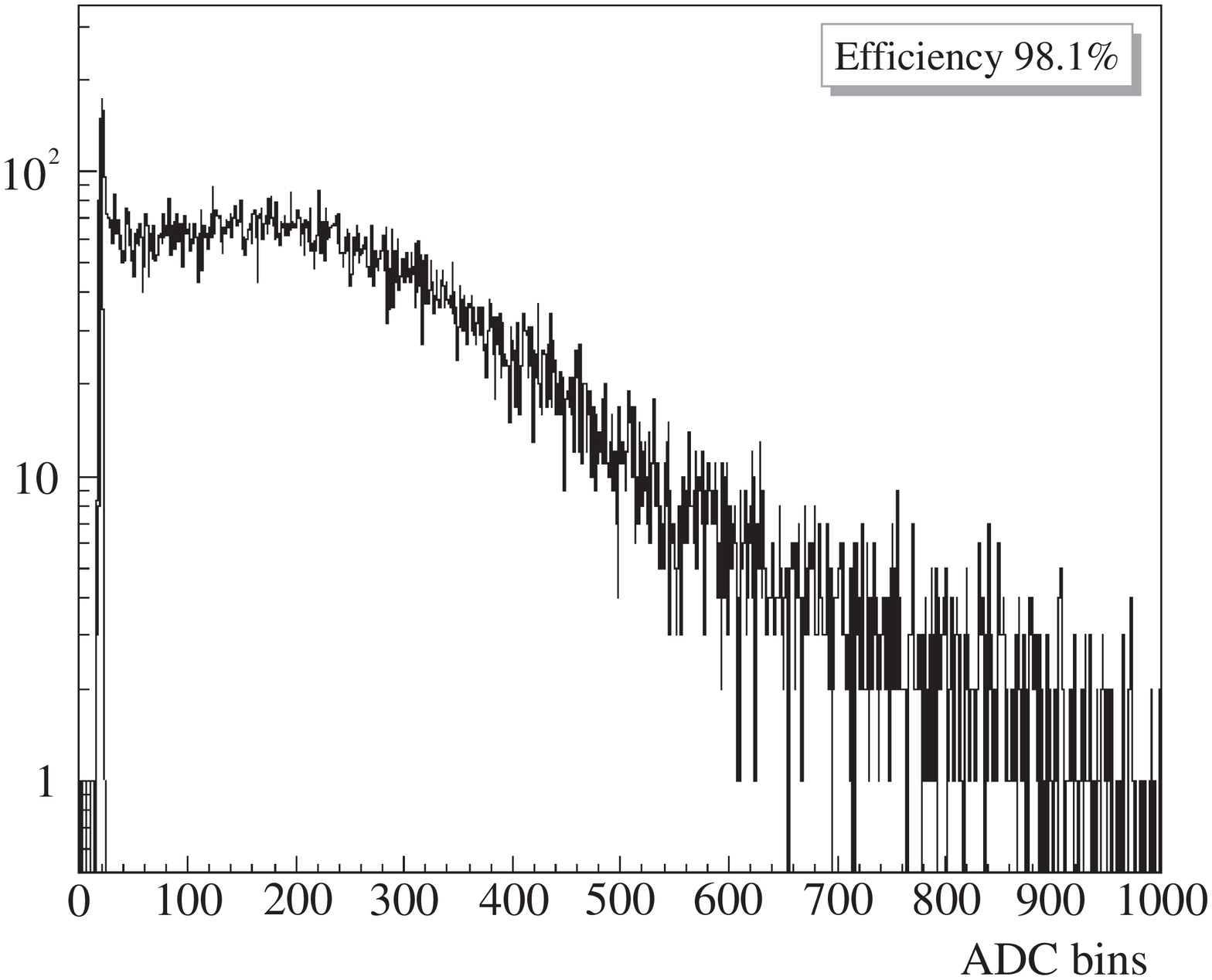} &
\includegraphics[width=.45\linewidth]{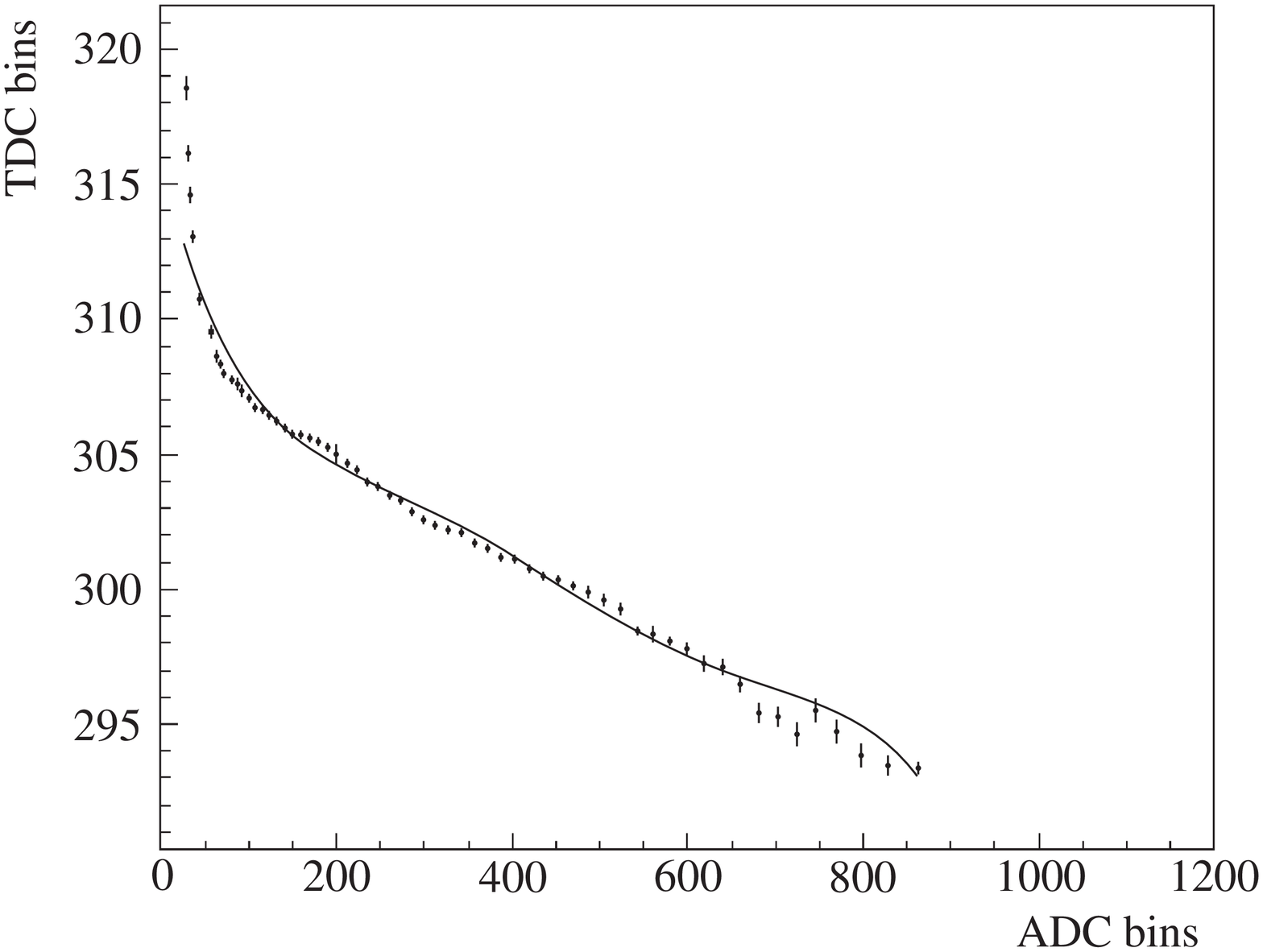} \\
{\large \it a} & {\large \it b} \\
\includegraphics[width=.45\linewidth]{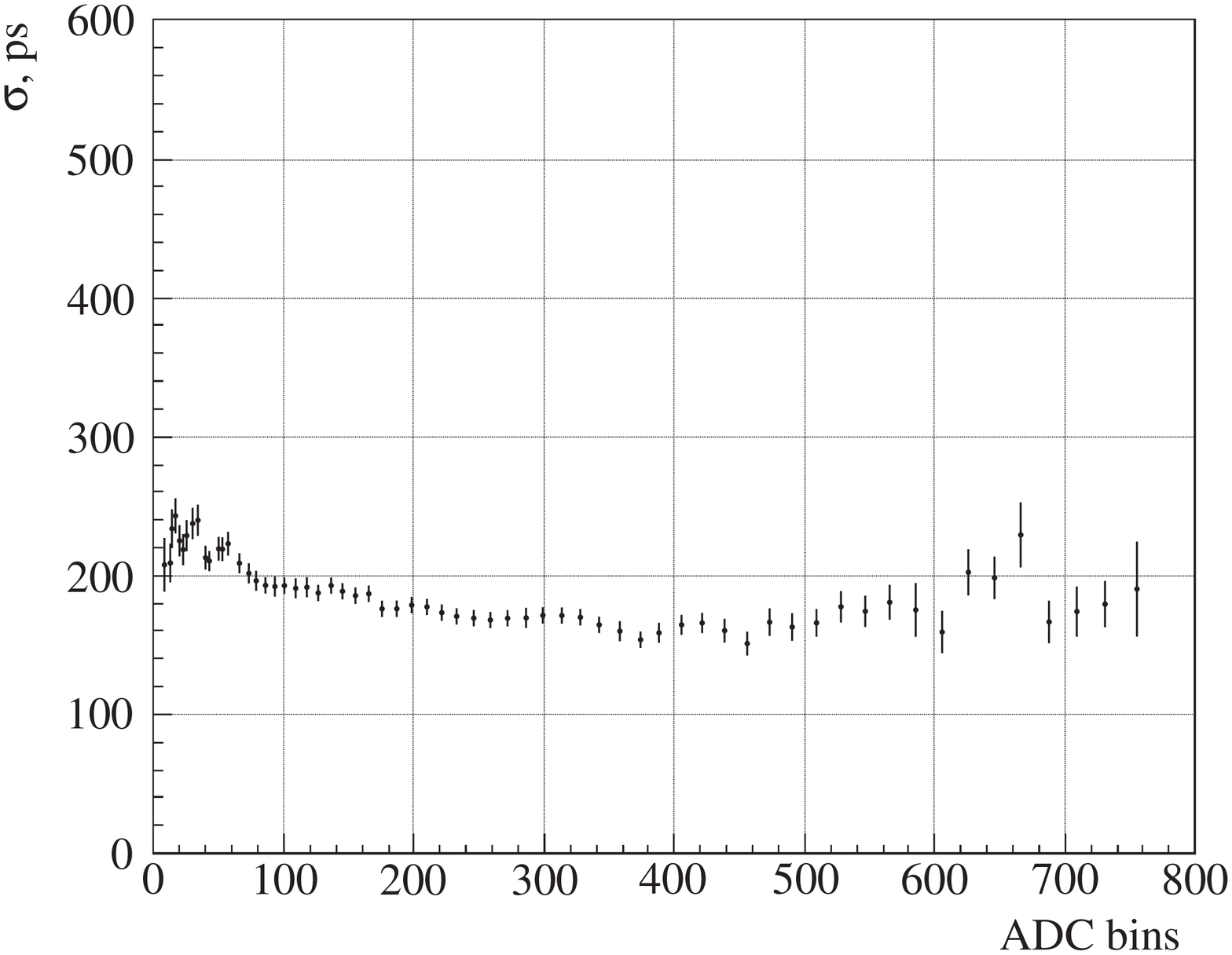} &
\includegraphics[width=.45\linewidth]{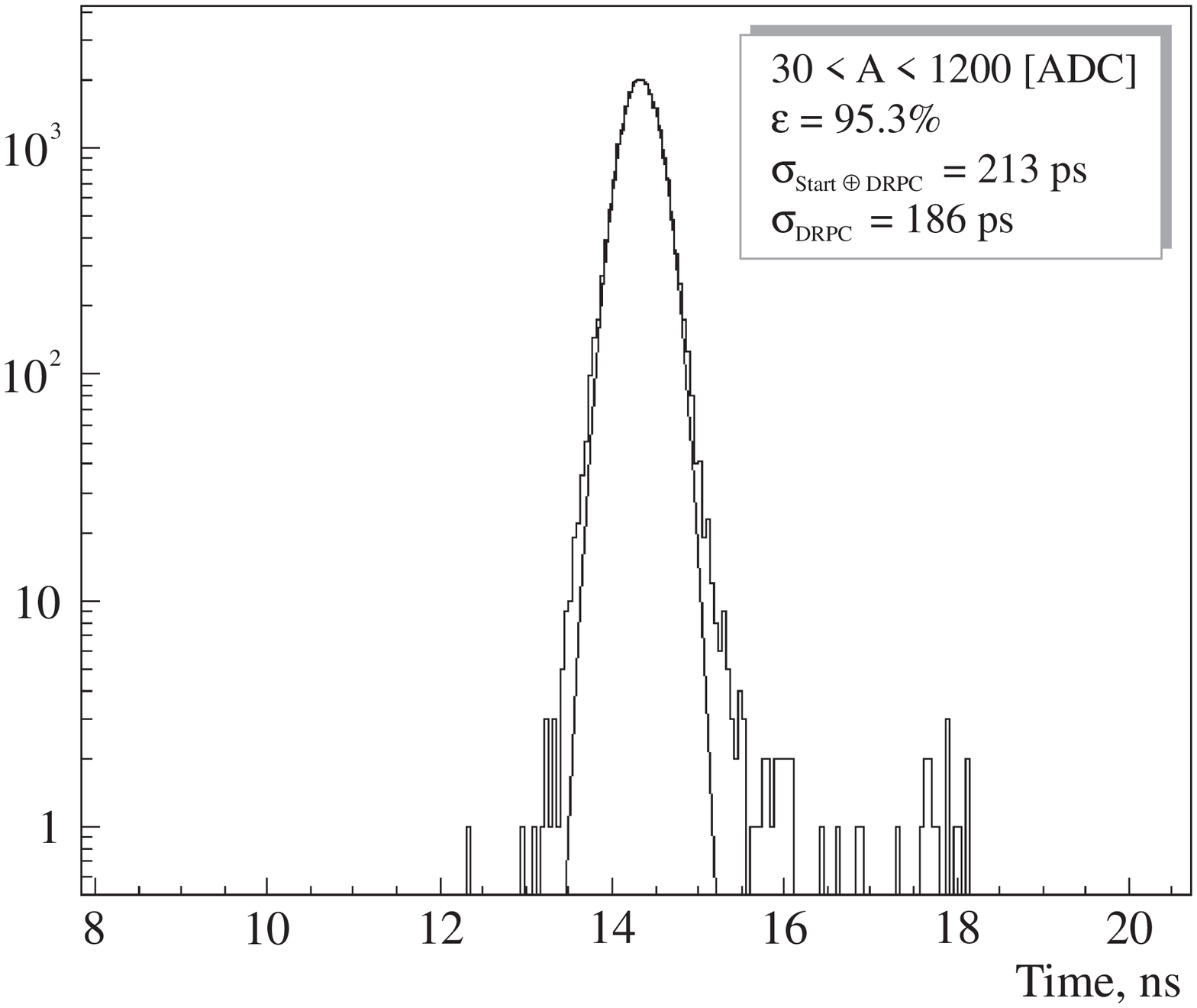} \\
{\large \it c} & {\large \it d} \\
\end{tabular}
\end{center}
\caption{Sample result of the DRPC time resolution and efficiency
  measurements: {\it a}~--- amplitude spectrum, {\it b}~--- polynomial
  T(A) correction, {\it c}~--- time resolution at different
  amplitudes, {\it d}~--- total time resolution.} 
\label{sevent}
\end{figure}

Fig.~\ref{sevent} represents a typical result of the DRPC
efficiency and time resolution measurements.

Plot in Fig.~\ref{sevent}{\it a} shows the amplitude distribution
measured with ADC. A very narrow pedestal, corresponding to noise
characterictics of both the detector and the electronic channel, may
be clearly seen. Efficient signals have a wide non-exponential
distribution. Such a shape allows to achieve high efficiency 
(98\% in this particular case) even if the detecting threshold 
of the electronics is high.

Distribution of {\it start} minus {\it stop} time versus 
amplitudes is shown in Fig.~\ref{sevent}{\it b}. One can clearly 
see that the mean time changes with the amplitudes for about 1~ns 
(this value must correspond to the signal rise time), since a 
constant threshold discriminator produces different times for 
jitters with different amplitudes. The solid line shows a 
polynomial approximation of this correlation. 

Taking into account T(A) dependence leads to the time resolution 
being dependent on amplitudes in the way presented in
Fig.~\ref{sevent}{\it c}. It may be seen that the time resolution
is close or even better than 200~ps in a wide range of apmlitudes,
excluding the lower region close to the discriminator threshold of
15~mV.

Finally, the DRPC time resolution at a fixed efficiency is shown in
Fig.~\ref{sevent}{\it d}. Better than 200~ps time resolution may
be reached at a high registration efficiency (95\% in this case). 

\begin{figure}
\begin{center}
\begin{tabular}{ccc}
\includegraphics[width=.3\linewidth]{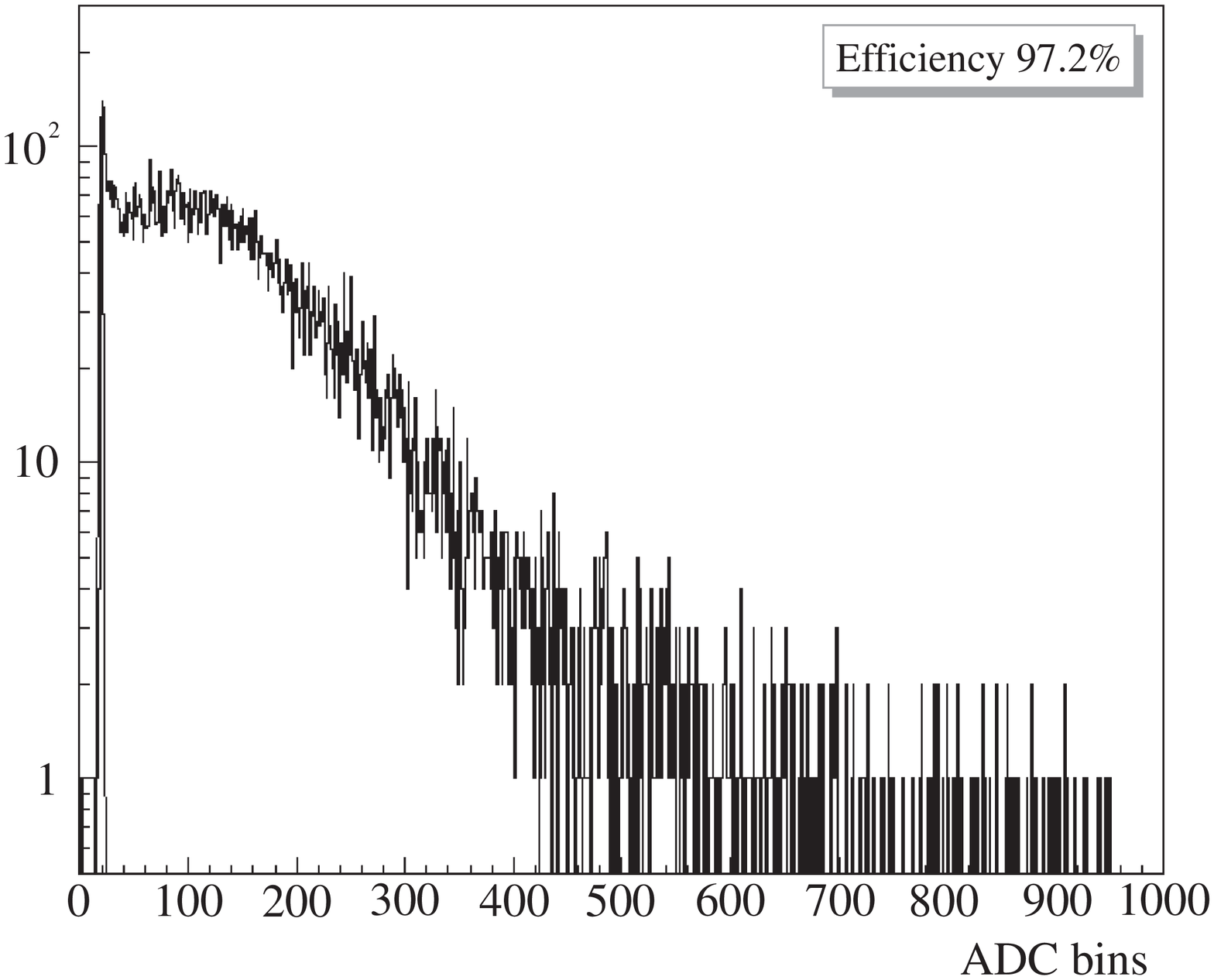} &
\includegraphics[width=.3\linewidth]{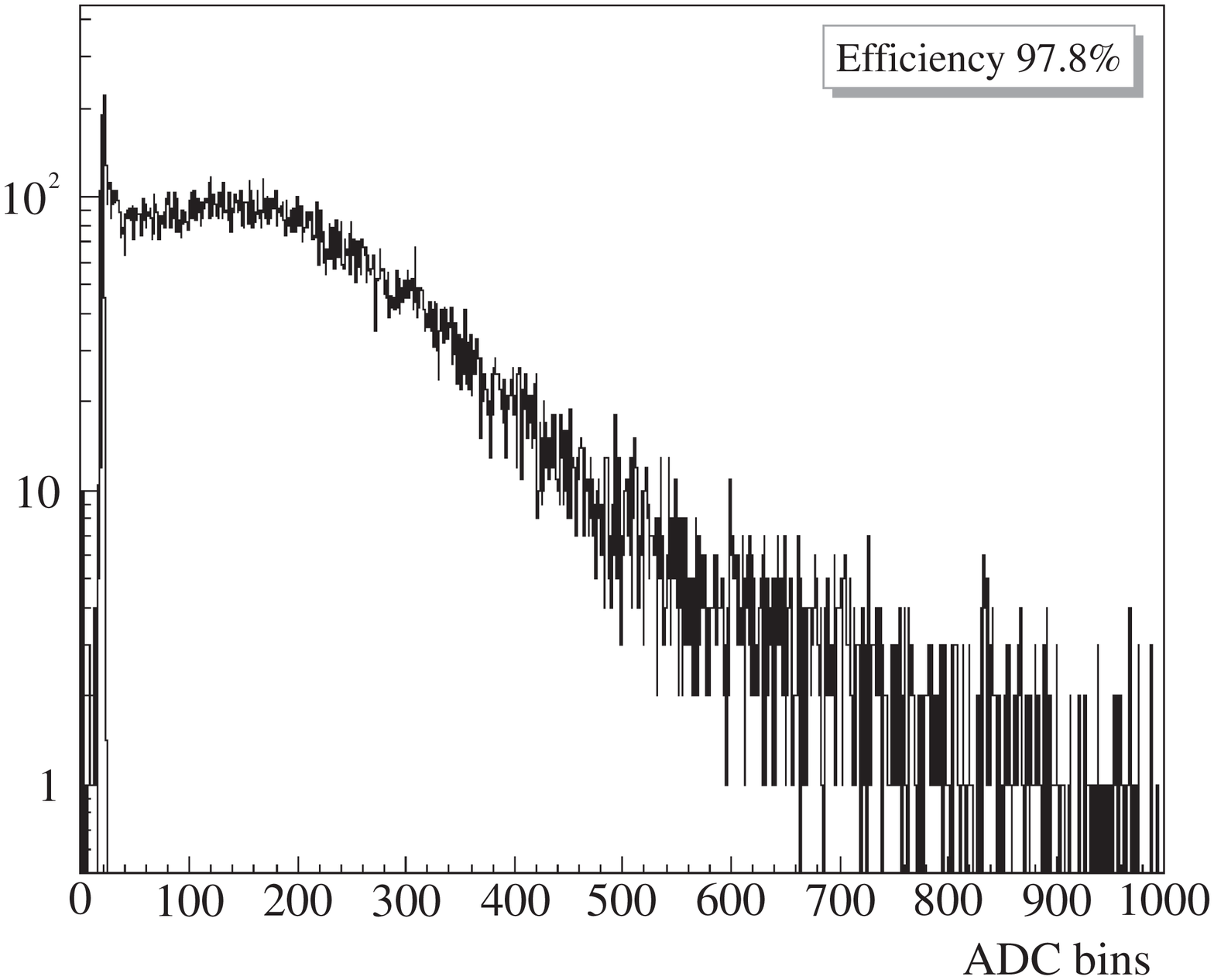} &
\includegraphics[width=.3\linewidth]{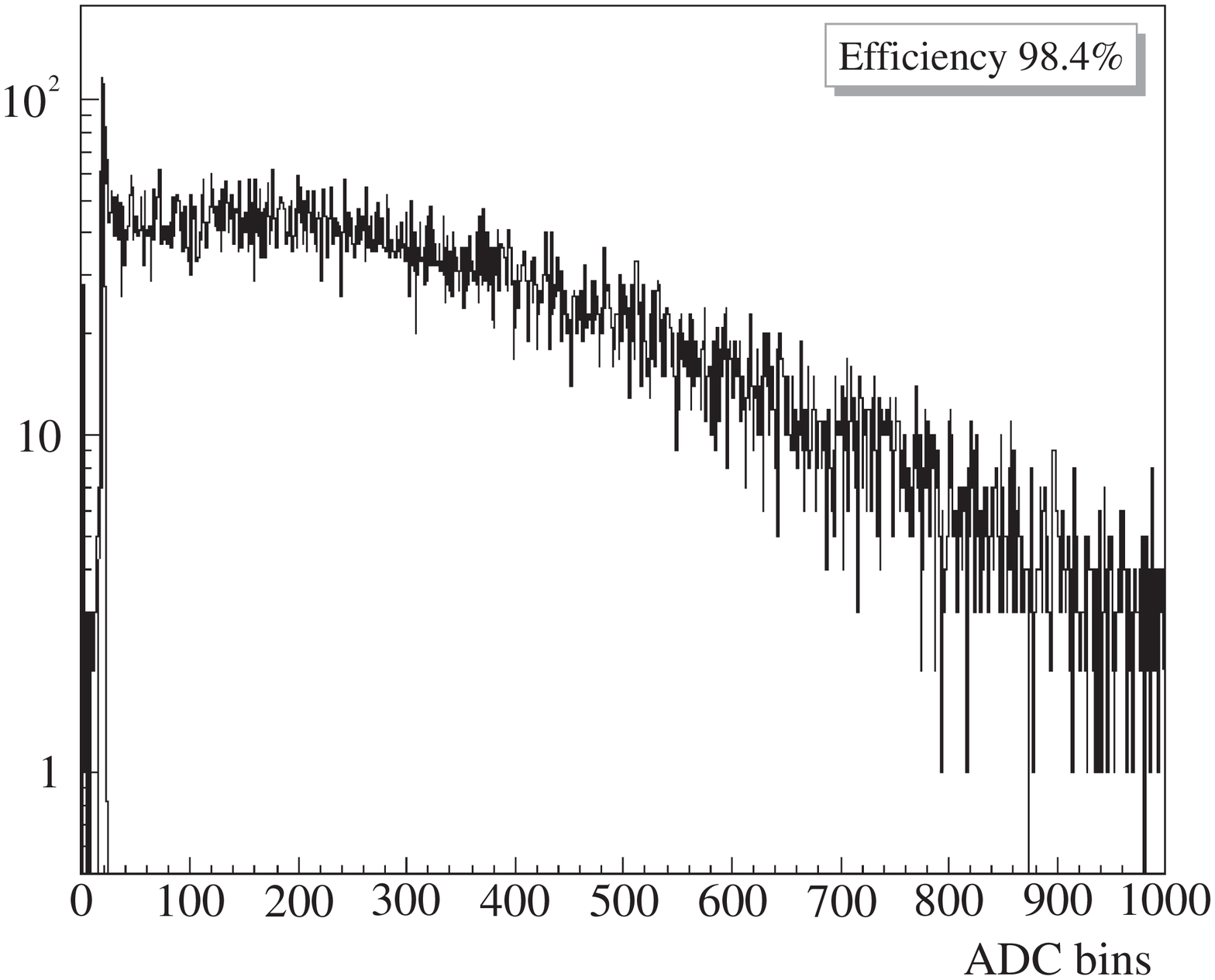} \\
\includegraphics[width=.3\linewidth]{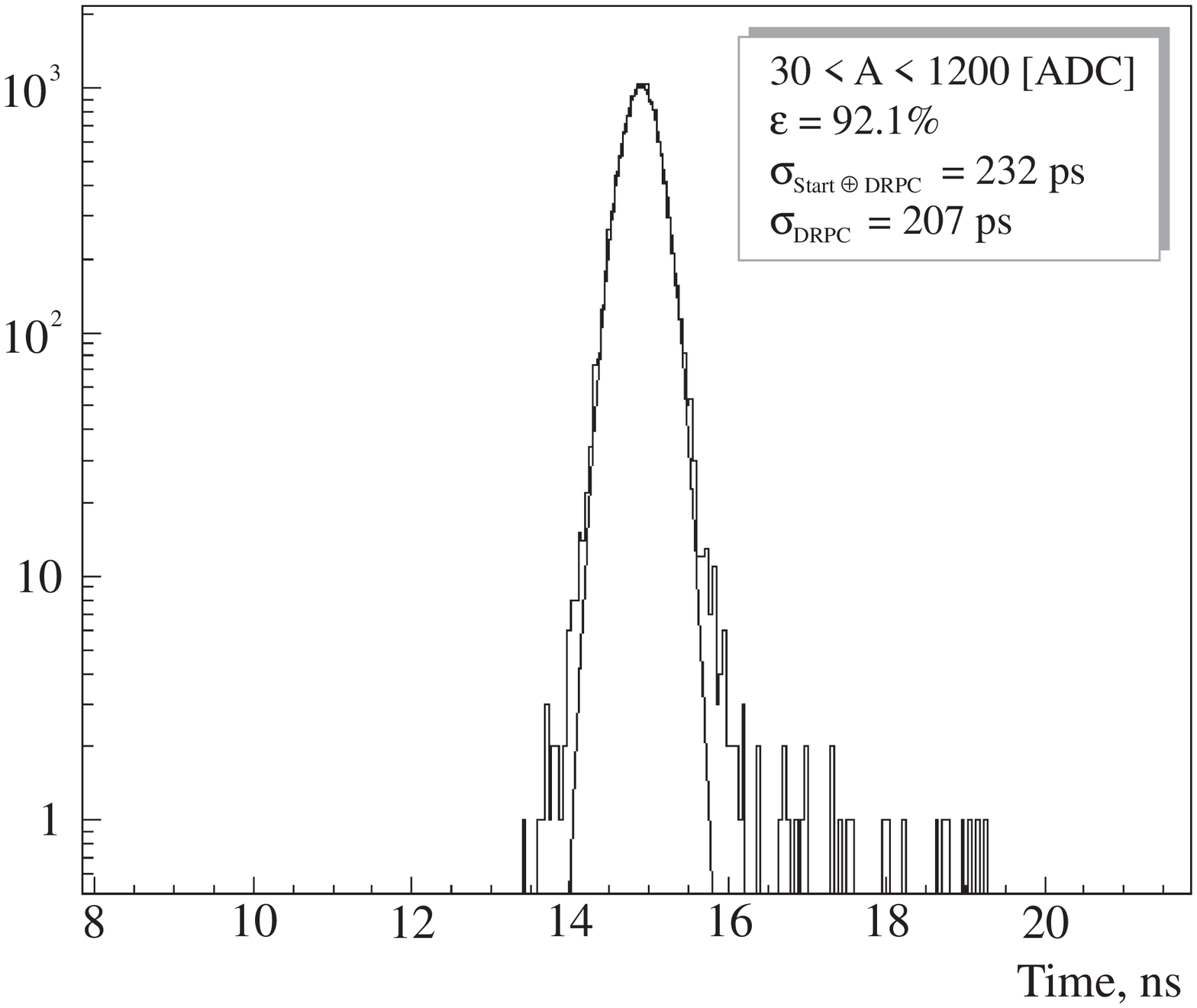} &
\includegraphics[width=.3\linewidth]{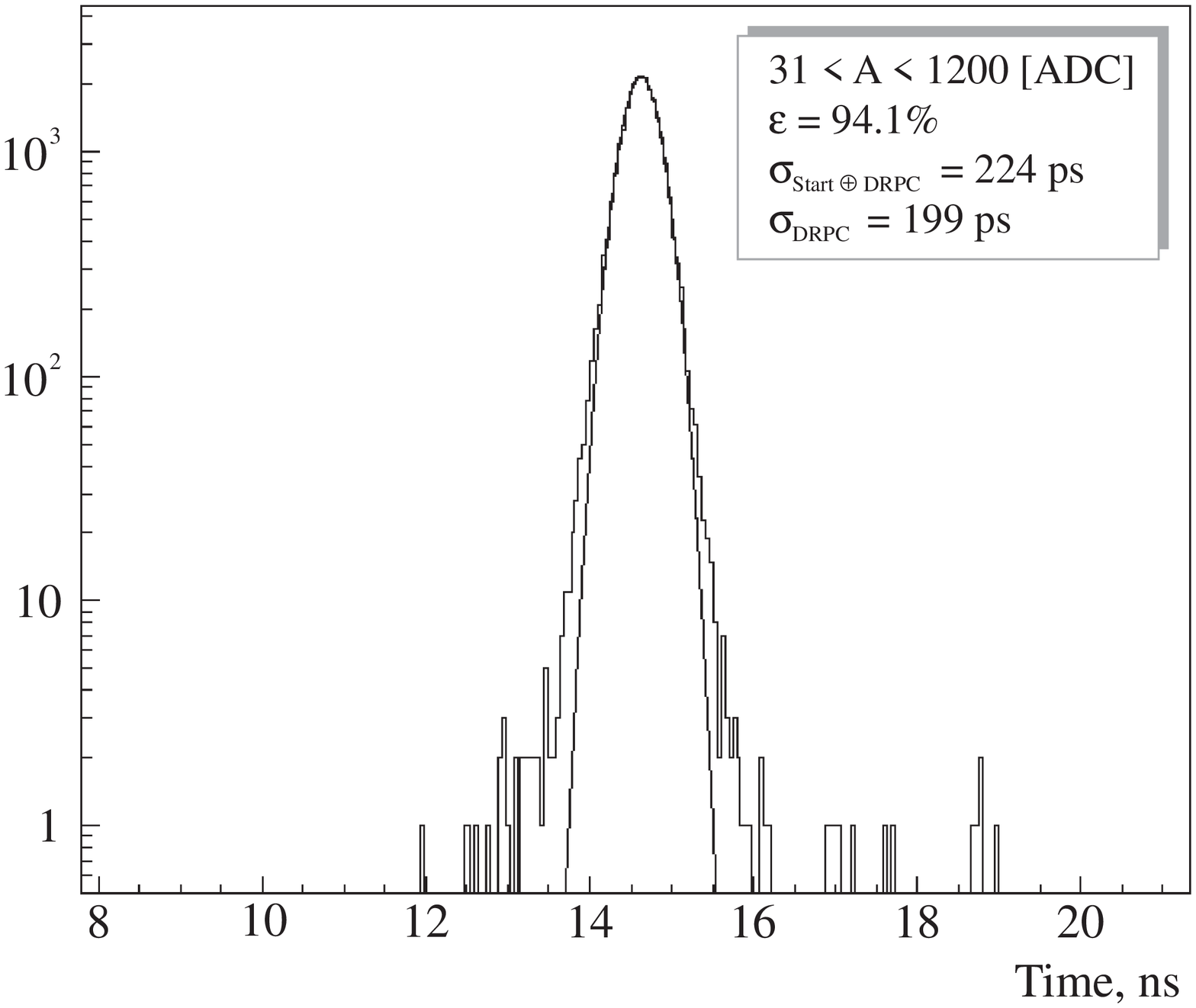} &
\includegraphics[width=.3\linewidth]{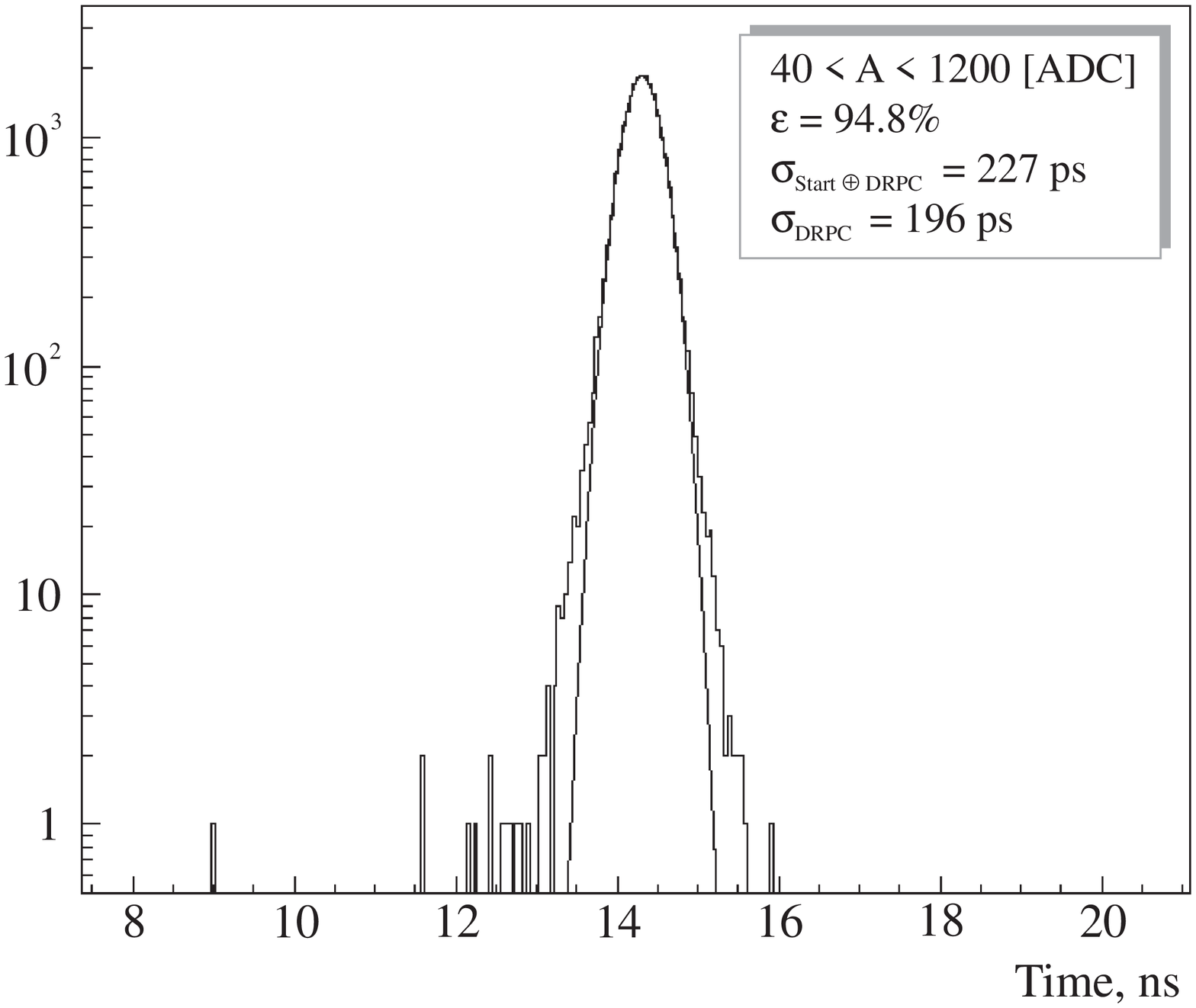} \\ \\ \\
\includegraphics[width=.3\linewidth]{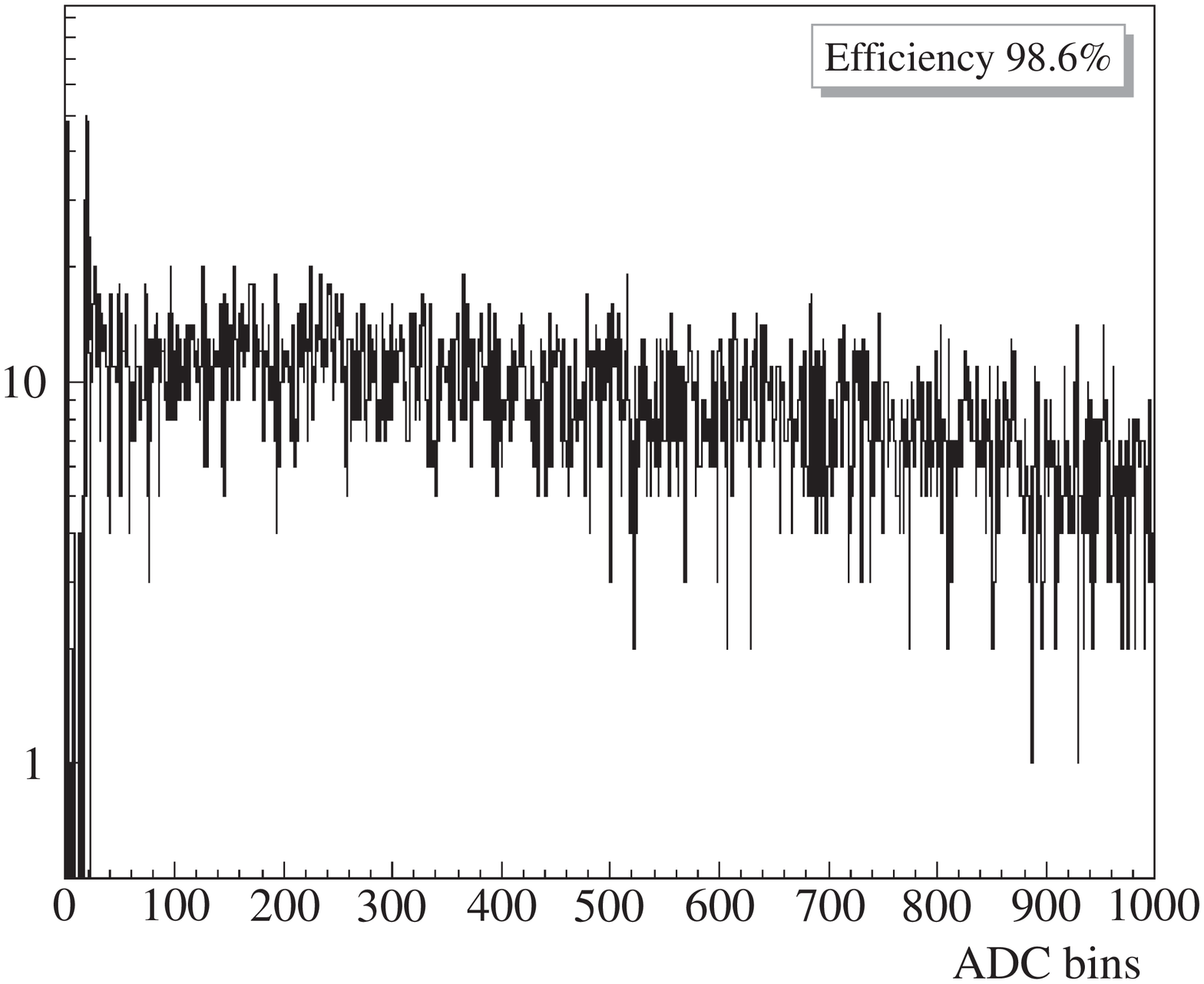} &
\includegraphics[width=.3\linewidth]{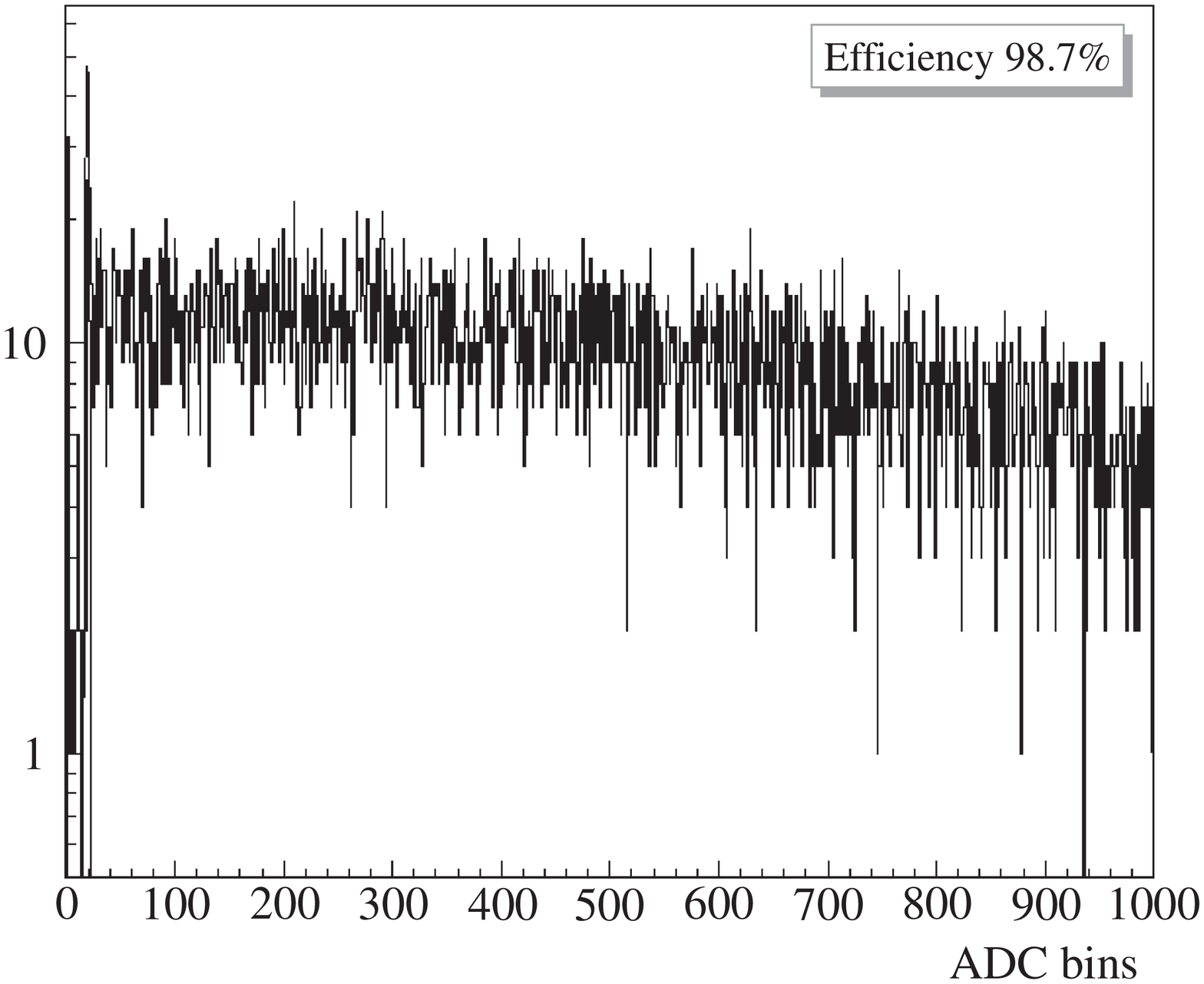} &
\includegraphics[width=.3\linewidth]{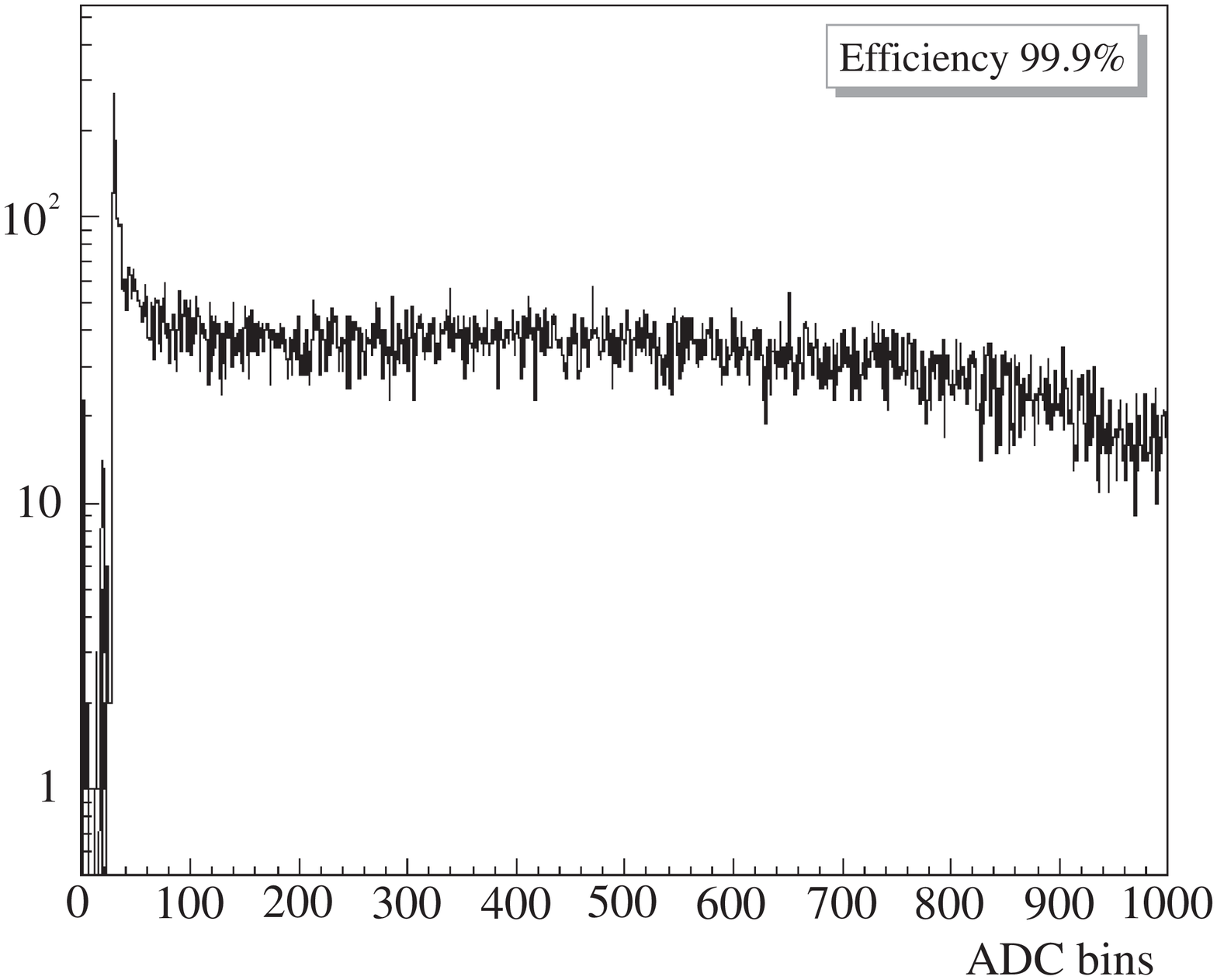} \\
\includegraphics[width=.3\linewidth]{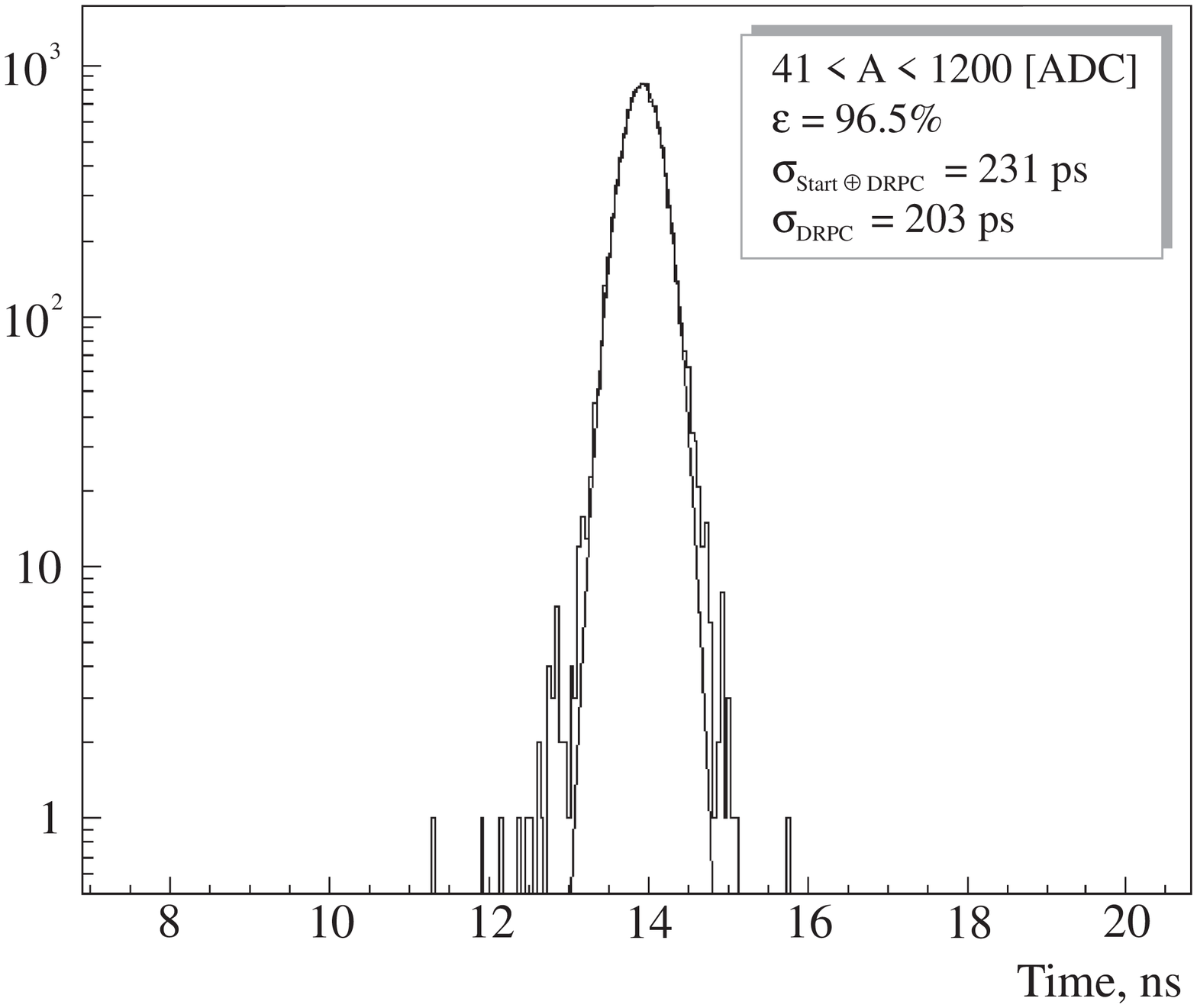} &
\includegraphics[width=.3\linewidth]{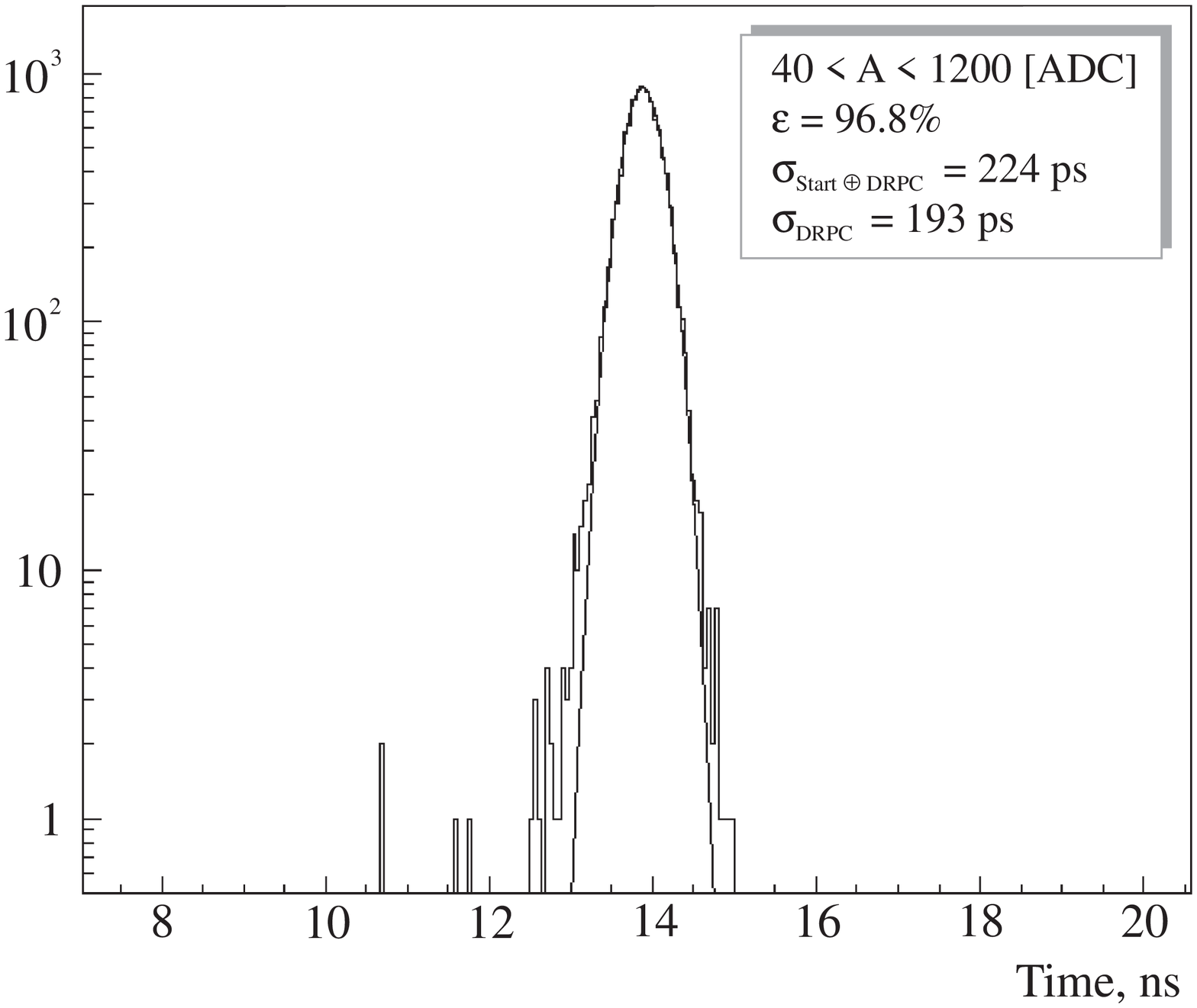} &
\includegraphics[width=.3\linewidth]{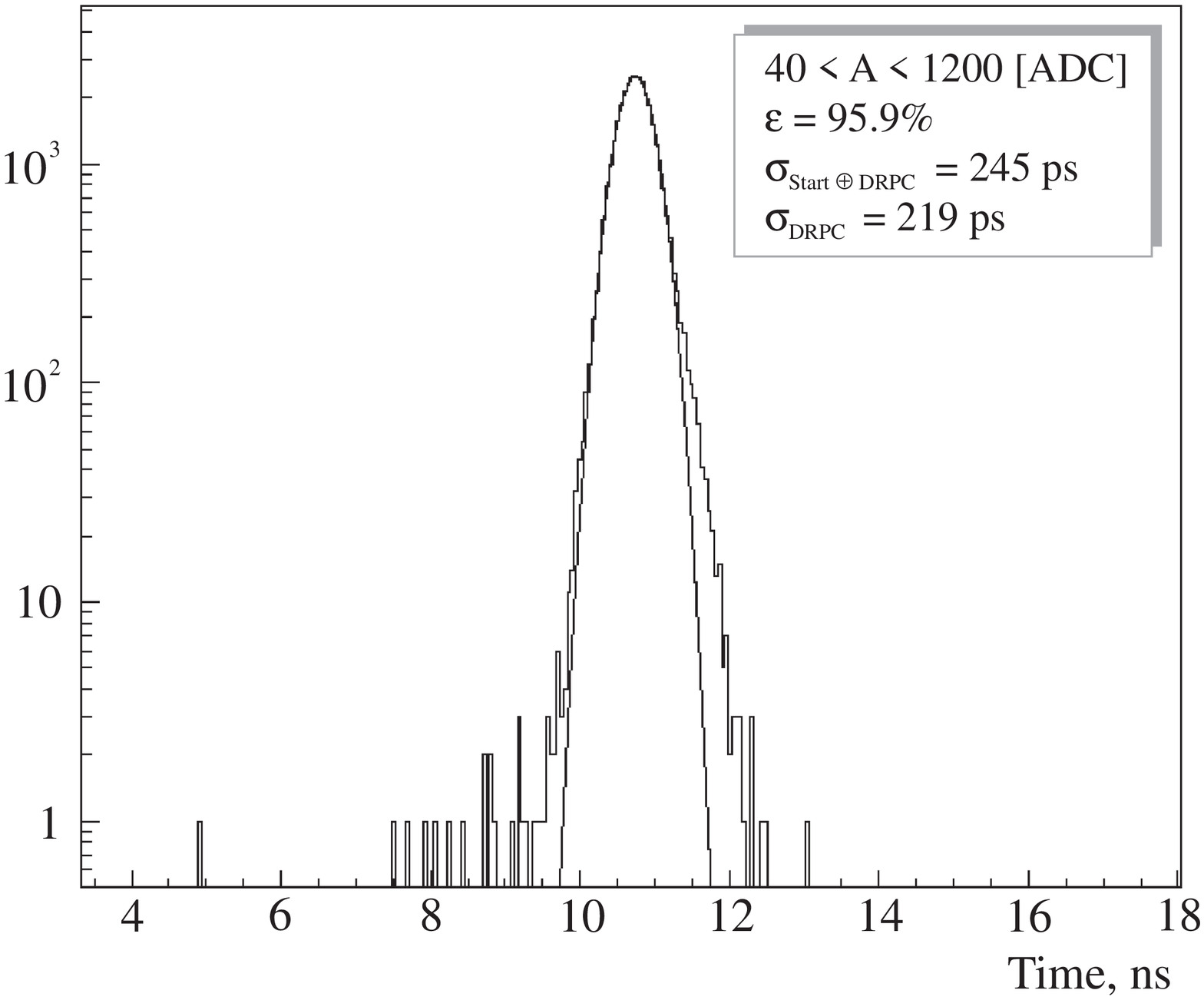} \\
\end{tabular}
\end{center}
\caption{Result of the time resolution and efficiency measurements at
  different applied voltages (100~V high voltage increment from plot
  to plot).}
\label{many}
\end{figure}

All the data shown in Fig.~\ref{sevent} were obtained at a 
fixed high voltage applied to the chamber. The fact that the 
time resolution does not depend on the applied voltage in its 
wide range is demonstrated in Fig.~\ref{many}. Amplitude 
distributions and timing characteristics
are plotted for several values of applied voltages starting from
3.7~kV with 100~V increment from plot to plot. Time resolution 
remains almost the same at different voltages (0.5~kV plateau) and
approximately equals to 200~ps. The detector efficiency stays 
high as well. 

DRPC time resolution of 200~ps is close to the same value 
measured for PPC with the 0.6~mm gas gap. It has been seen that 
the PPC time resolution decreases approximately in a linear way 
as the gap becomes narrower, and 0.6~mm is a minimal value 
still giving high efficiency of MIP registration with double-gap 
detectors. In this sence, one can expect that, similar to PPC, 
the DRPC time resolution of 200~ps may be significantly improved 
by decreasing the gap width.

\section{Limited energy resolution in breakdowns}

It has been known so far that one of the main disadvantages 
of PPC is its high energy resolution during spontaneous 
breakdowns, which is limited only by energy acquired during the 
discharge. Despite the very low probability of such a breakdown 
(10$^{-6}$ if modern freon-based gases are used), the energy 
resolution is about 6 orders of magnitude higher than in 
ordinal avalanche cases. Besides harming the detector and 
electronics, such a huge energy resolution leads to cross-talks 
between different cells of the big module and to a large 
dead time of the detector after breakdowns. 

DRPC is suggested as a principal solution in the task of suppressing
the energy resolution during breakdowns. 

\begin{figure}
\begin{center}
\begin{tabular}{cc}
\includegraphics[width=.45\linewidth]{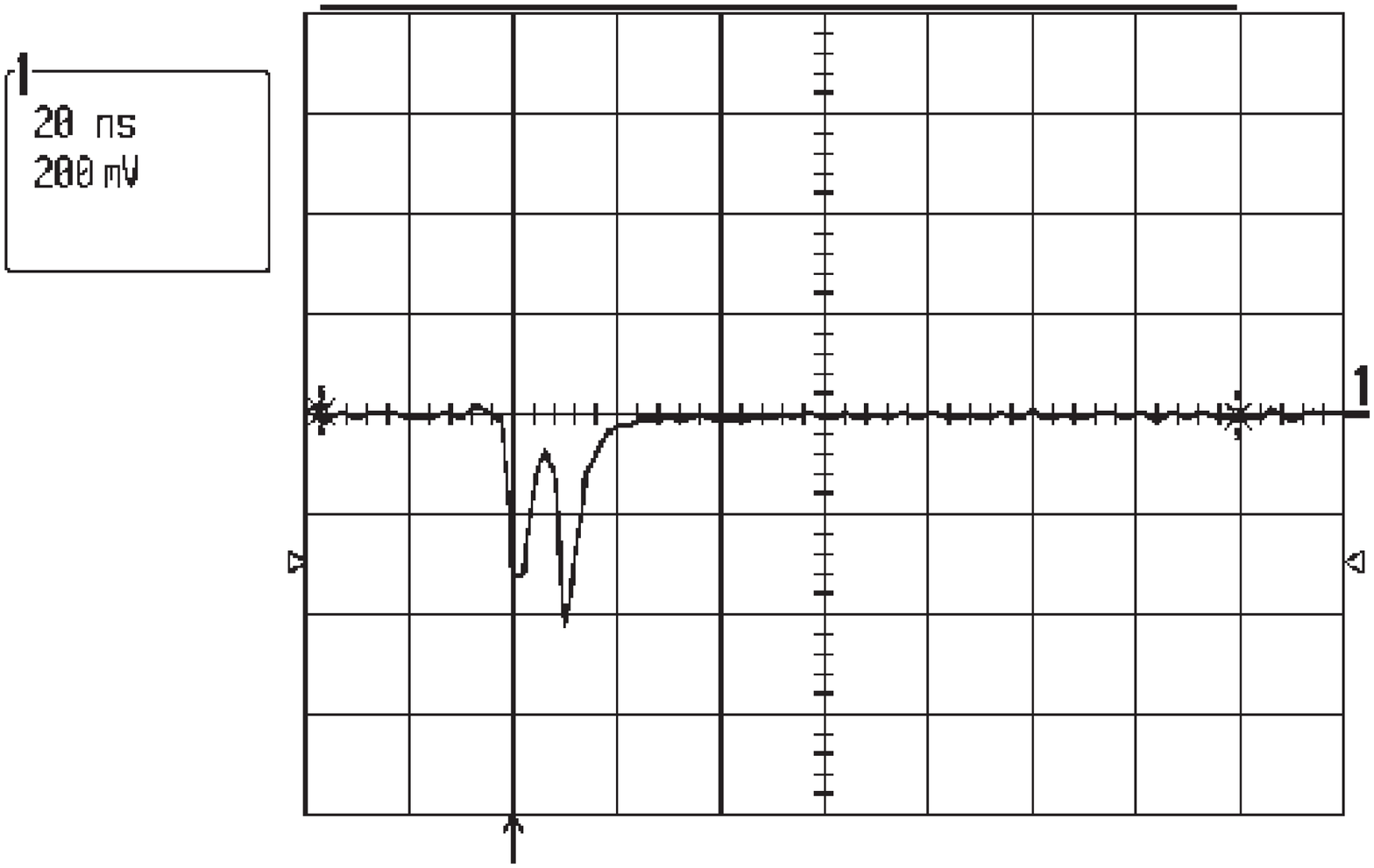} &
\includegraphics[width=.45\linewidth]{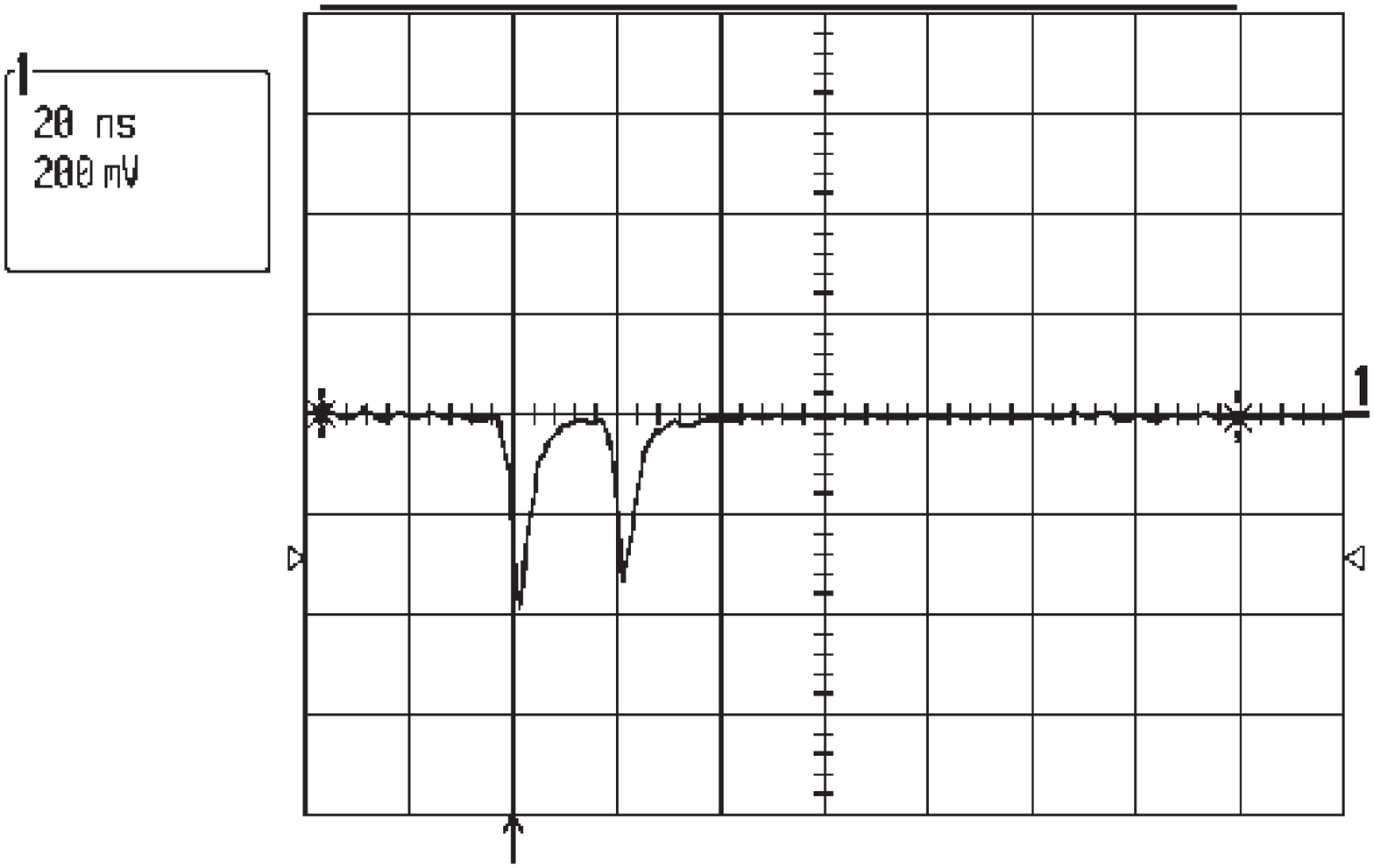} \\
\includegraphics[width=.45\linewidth]{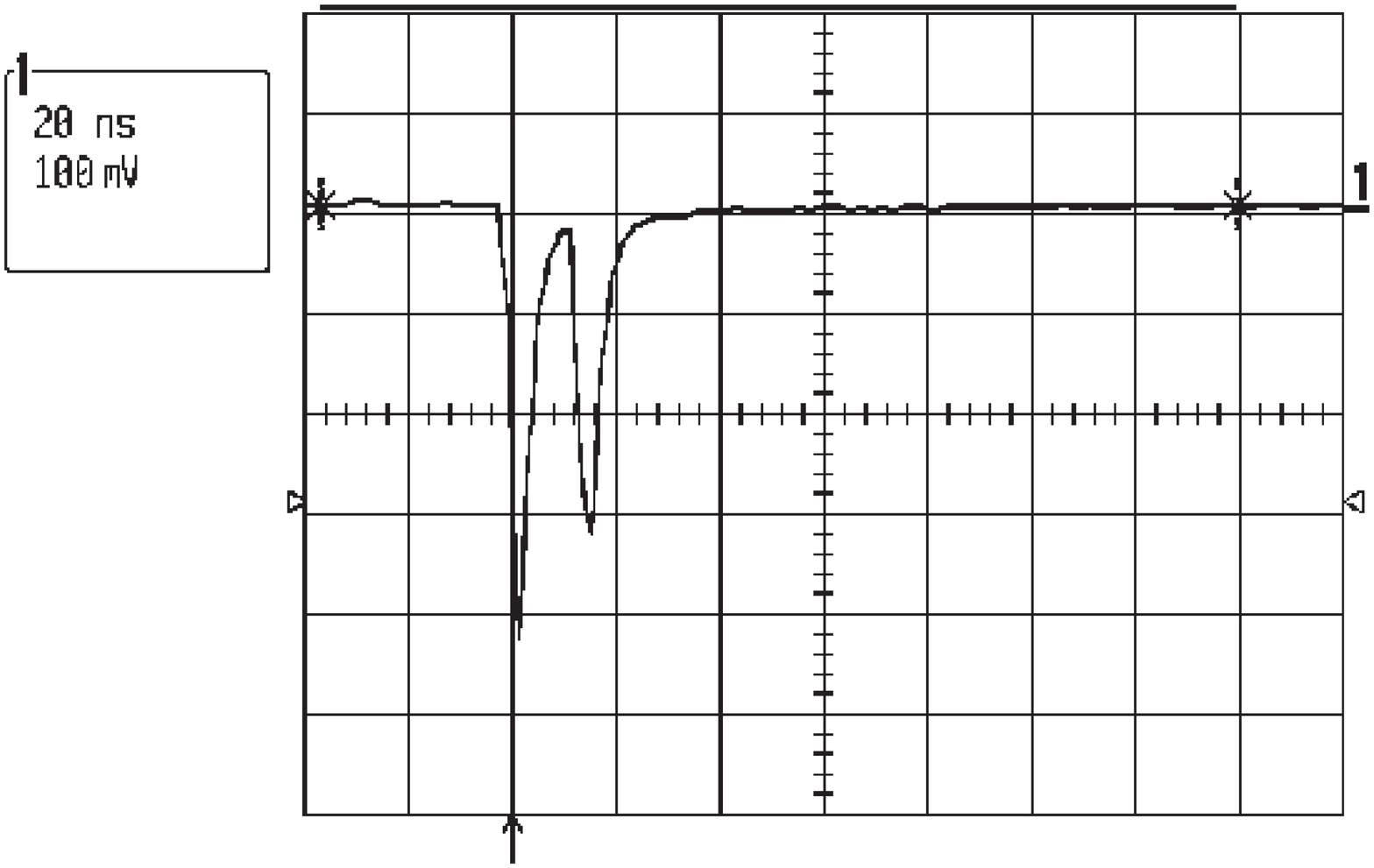} &
\includegraphics[width=.45\linewidth]{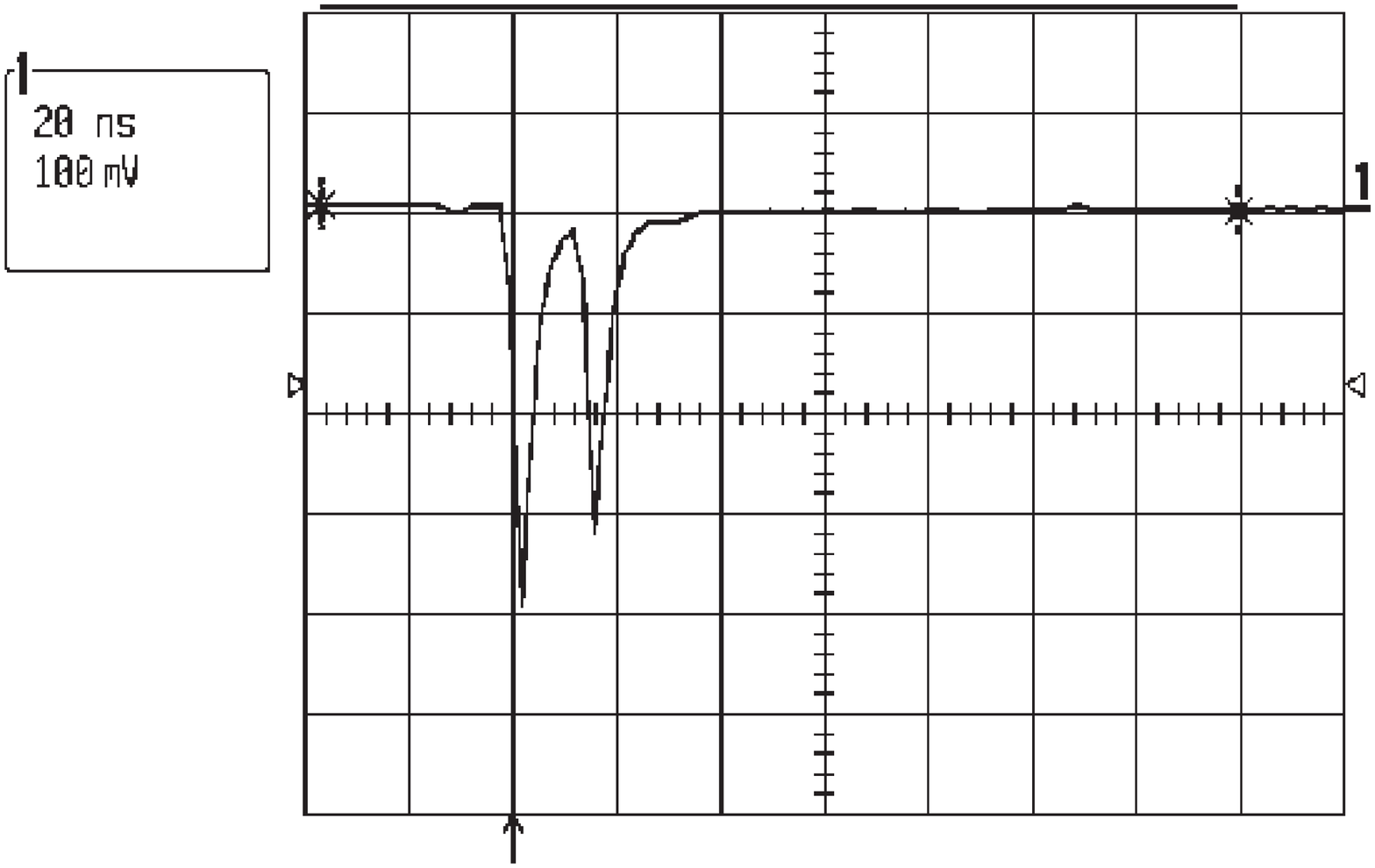} \\
\end{tabular}
\end{center}
\caption{Oscilograms of breakdown signals measured directly from the
  detector with 50~$\Omega$ output resistor.} 
\label{direct}
\end{figure}

Oscillograms in Fig~\ref{direct} show breakdown signals from DRPC
without further amplification and with 50~$\Omega$ 
resistor at the output. Signal shapes have two bumps 
(seen with PPC as well), very short total duration of 
about 20~ns, and small amplitudes of 100--200~mV. Summarized 
energy resolution is about 1~nC which is 2
orders of magnitude lower than in the PPC case.  

\begin{figure}
\begin{center}
\begin{tabular}{cc}
\includegraphics[width=.45\linewidth]{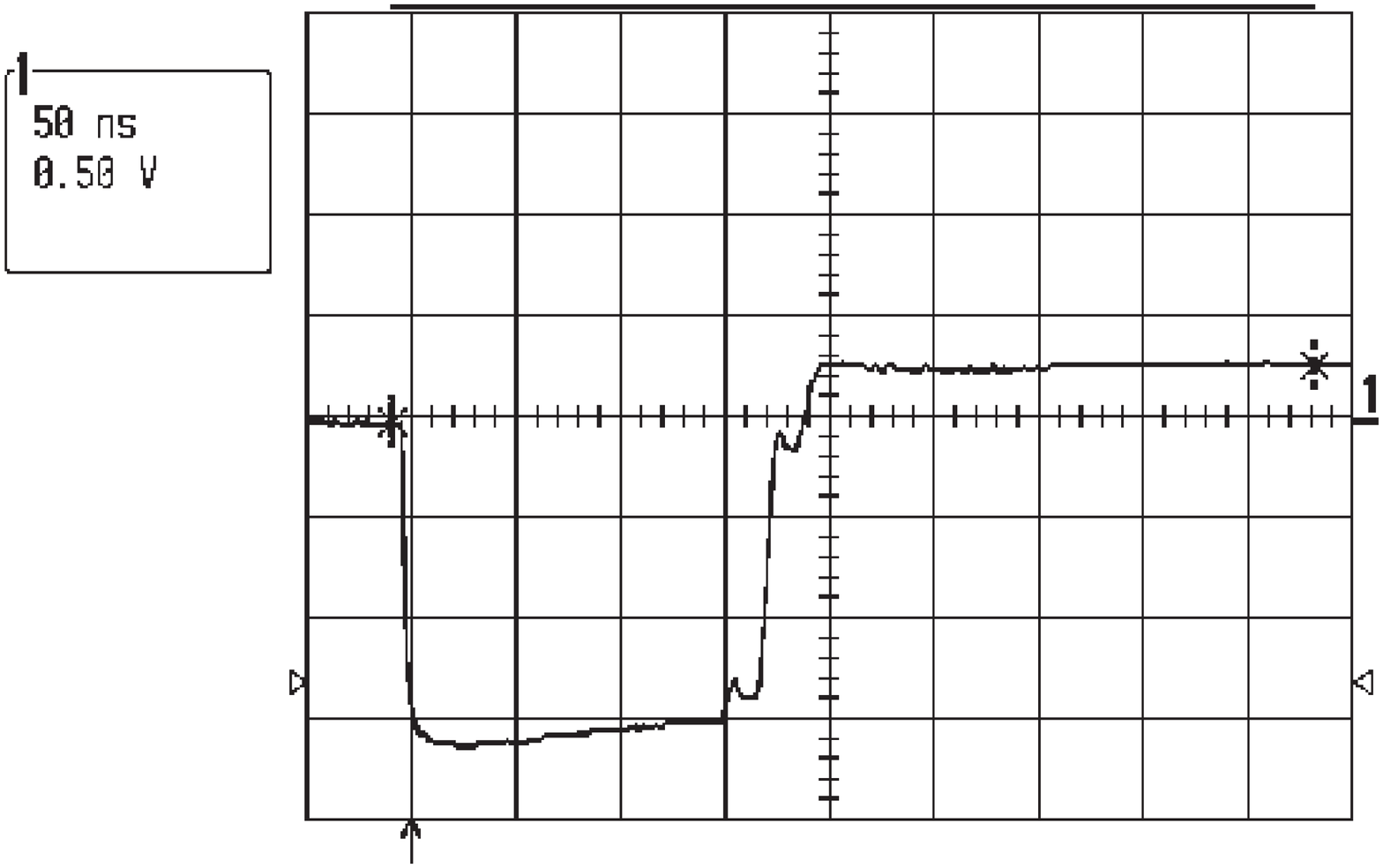} &
\includegraphics[width=.45\linewidth]{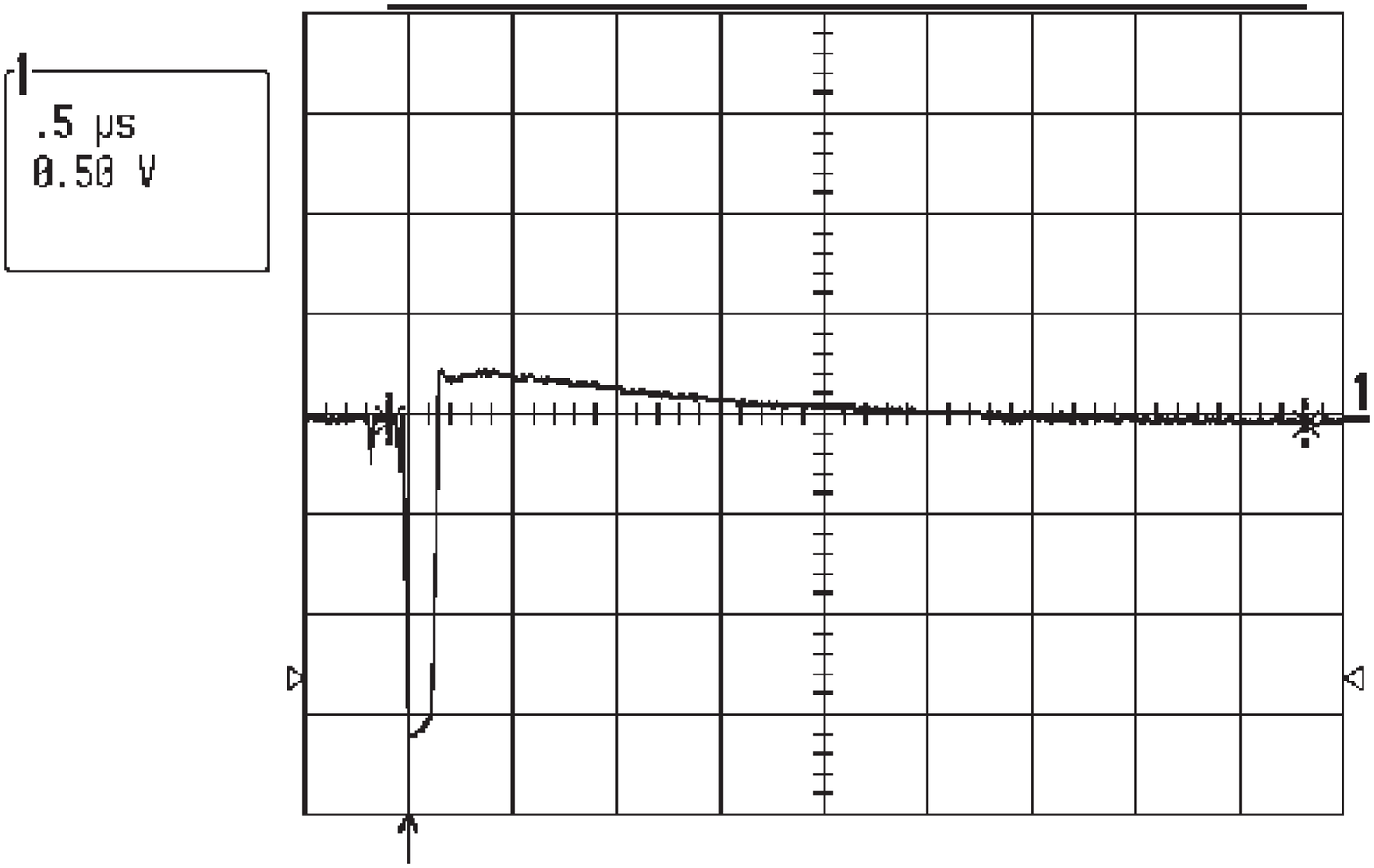} \\
\end{tabular}
\end{center}
\caption{Oscillograms of fast electronics responce for breakdown signals.}
\label{electr}
\end{figure}

Fig.~\ref{electr} represents the reaction of fast sensitive
electronics on such a limited breakdown signal. One can see 
that the amplifier becomes saturated, but only for a short 
time period of about 200~ns, which cannot influence the total 
dead time of the detector. 

\section{Admixture of streamer signals}

As in the PPC case, the avalanche evolution inside DRPC may 
lead to a streamer formation. The streamer probability depends 
primarily on the chosen gas, and may be either strongly 
suppressed, or released to let
the detector work in the streamer mode. 

We were not aimed in choosing or optimising working regime of 
DRPC but have seen a small admixture of streamer signals, 
rising to a 10$^{-2}$ level at the end of the counting plateau, 
with DME + CF$_3$Br as a working gas. 

\begin{figure}
\begin{center}
\begin{tabular}{cc}
\includegraphics[width=.45\linewidth]{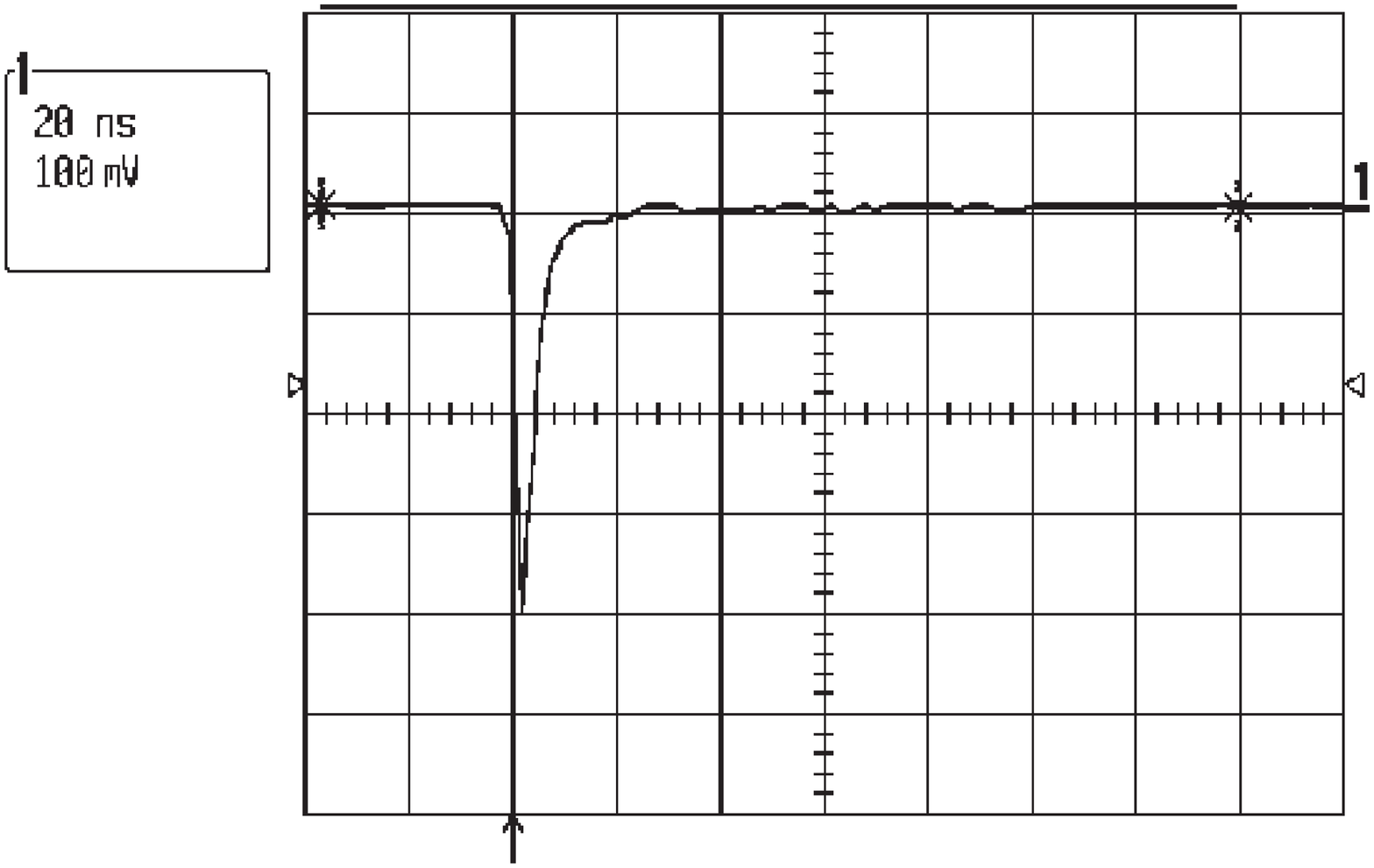} &
\includegraphics[width=.45\linewidth]{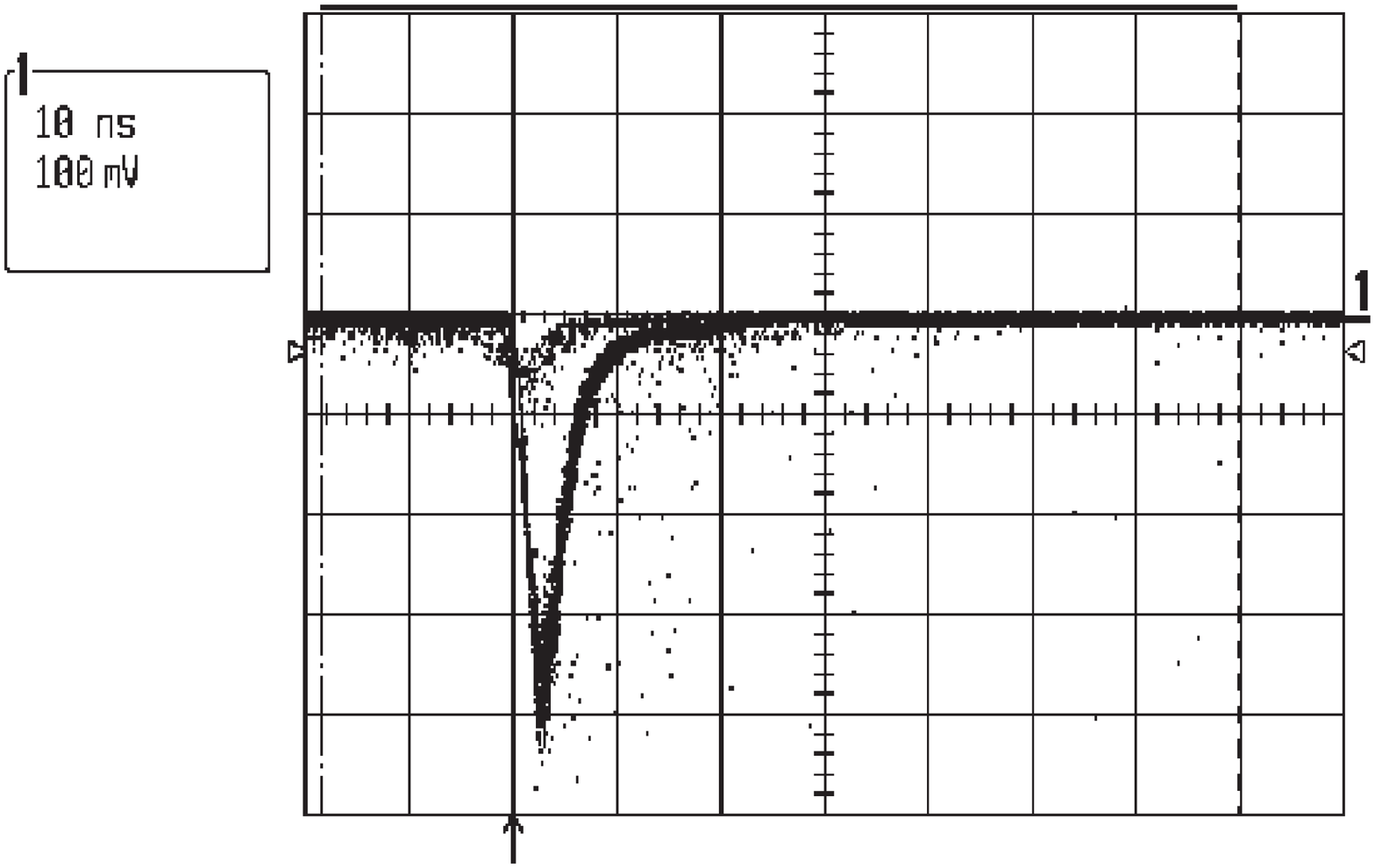} \\
{\large \it a} & {\large \it b} \\
\end{tabular}
\end{center}
\caption{Oscillograms of a single streamer signal ({\it a}) and
mixture of streamer and avalanche signals at the upper region of the
counting plateau ({\it b}).}
\label{streamer}
\end{figure}

Oscillograms in Fig.~\ref{streamer} show the single streamer signal
from DRPC ({\it a}), and mixture of avalanche and streamer signals
obtained at the upper region of the counting plateau ({\it b}) 
without amplification. 

All streamer signals have same shapes with fast rise times and large
amplitudes. Such working regime seems to be very attractive and must
be additionally investigated. 

\section{Cost estimations of a large TOF system based on DRPCs}

The cost of a large TOF system built of many DRPC cells may be
estimated supposed that the single channel price is about
\$40--50, including the cost of development of new front-end
and readout electronics,
prototype construction, etc. This price is more than an order of
magnitude less, than that which might be required for TOF based on
scintillation counter technique. 

Moreover, it is obvious that timing measurements are not the only
possible employment of such a simple and cheap detector.

\end{document}